\documentclass[pdftex,letterpaper,12pt,authoryear,final]{elsarticle}

\usepackage{graphicx}
\usepackage{natbib}
\usepackage{amssymb}
\usepackage{amsmath}
\usepackage{lineno}
\usepackage{url}
\usepackage{longtable}

\usepackage{hyperref}

\usepackage{textcomp}

\usepackage{fancyhdr}
\bibpunct[ ]{(}{)}{,}{a}{}{,}

\newif\ifincludegraphics
\includegraphicstrue

\newenvironment{SmallLongTable}{\begin{footnotesize}}{\end{footnotesize}}

\usepackage[width=6.5in,height=9in]{geometry}

\newcommand\micron{\mbox{\textmu{}m}}%
\newcommand\arcdeg{\mbox{$^\circ$}}%
\newcommand\arcsec{\mbox{$^{\prime\prime}$}}%
\newcommand\degr{\arcdeg}%
\newcommand\sun{\odot}%
\newcommand\mH{\mbox{$m_{\rm H}$}} 
\newcommand\coo{CO$_2$}
\newcommand\water{\mbox{H$_2$O}}
\newcommand\di{\textit{Deep Impact}}
\newcommand\nodata{$\cdots$}%

\newcommand\fnorm{\mbox{$F_{\lambda,700}$}}
\newcommand\fmin{\mbox{$F_{\lambda,{\rm min}}$}}
\newcommand\fmax{\mbox{$F_{\lambda,{\rm max}}$}}

\newcommand\baas{\ref@jnl{BAAS}}%

\journal{Icarus}

\begin{document}
\begin{frontmatter}

\title{A distribution of large particles in the coma of Comet 103P/Hartley 2}

\author[umd]{Michael S. Kelley\corref{cor}}
\cortext[cor]{Corresponding author}
\ead{msk@astro.umd.edu}

\author[sigma]{Don J. Lindler}
\ead{don.j.lindler@nasa.gov}

\author[umd]{Dennis Bodewits}
\ead{dennis@astro.umd.edu}

\author[umd]{Michael F. A'Hearn}
\ead{ma@astro.umd.edu}

\author[apl]{Carey M. Lisse}
\ead{Carey.Lisse@jhuapl.edu}

\author[umd]{Ludmilla Kolokolova}
\ead{ludmilla@astro.umd.edu}

\author[mps,ret]{Jochen Kissel}

\author[uh]{Brendan Hermalyn}
\ead{hermalyn@ifa.hawaii.edu}

\address[umd]{Department of Astronomy, University of Maryland, College
  Park, MD 20742-2421, USA}

\address[sigma]{Sigma Space Corporation, 4600 Forbes Boulevard,
  Lanham, MD 20706, USA}

\address[apl]{Johns Hopkins University--Applied Physics Laboratory,
  11100 Johns Hopkins Road, Laurel, MD 20723, USA}

\address[mps]{Max-Planck-Institut f\"ur Sonnensystemforschung,
  Max-Planck-Str. 2, 37191 Katlenburg-Lindau, Germany}

\address[uh]{NASA Astrobiology Institute, Institute for Astronomy,
  University of Hawaii, 2680 Woodlawn Drive, Honolulu, HI 96822, USA}

\fntext[ret]{retired}

\begin{abstract}
The coma of Comet 103P/Hartley~2 has a significant population of large
particles observed as point sources in images taken by the \di{}
spacecraft.  We measure their spatial and flux distributions, and
attempt to constrain their composition.  The flux distribution of
these particles implies a very steep size distribution with power-law
slopes ranging from $-6.6$ to $-4.7$.  The radii of the particles
extend up to 20~cm, and perhaps up to 2~m, but their exact sizes
depend on their unknown light scattering properties.  We consider two
cases: bright icy material, and dark dusty material.  The icy case
better describes the particles if water sublimation from the particles
causes a significant rocket force, which we propose as the best method
to account for the observed spatial distribution.  Solar radiation is
a plausible alternative, but only if the particles are very low
density aggregates.  If we treat the particles as mini-nuclei, we
estimate they account for $<16-80$\% of the comet's total water
production rate (within 20.6~km).  Dark dusty particles, however, are
not favored based on mass arguments.  The water production rate from
bright icy particles is constrained with an upper limit of 0.1 to
0.5\% of the total water production rate of the comet.  If indeed icy
with a high albedo, these particles do not appear to account for the
comet's large water production rate.

\textbf{Erratum:} We have corrected the radii and masses of the large
particles of comet 103P/Hartley 2 and present revised conclusions in
the attached erratum.

\end{abstract}

\end{frontmatter}
\thispagestyle{fancy}
\section{Introduction}
Comet 103P/Hartley 2 (hereafter, 103P or Hartley 2) is a hyperactive
comet.  The activities of most comets are consistent with surfaces
that are over 90\% inert, i.e., activity is restricted to a few
localized sources.  In contrast, the gas production rates of the
hyperactive comets suggest activity over $\sim100$\% of their surfaces
or more.  Comet Hartley 2's peak water production rate during the 1997
perihelion passage was $Q_{H_2O} \approx 3\times10^{28}$
molec~s$^{-1}$ \citep{combi11-h2}.  The sublimation rate of an
isothermal nucleus with a 4\% Bond albedo at 1.03~AU (the 1997
perihelion distance of Hartley 2) is $2.9\times10^{17}$
molec~cm$^2$~s$^{-1}$ \citep{cowan79}.  Thus, approximately 10~km$^2$
of surface area would be required to match the measured water
production in 1997.  However, constraints on the size of the nucleus
by \citet{groussin04-comets} and \citet{lisse09} suggested the total
surface area was near 4--6~km$^2$, yielding an active fraction (area
based on $Q_{H_2O}$/area of the nucleus) near 1.7--2.5.  In order to
account for the high active fraction, \citet{lisse09} proposed that
the coma contained a population of icy grains, which increased the
surface area available for sublimation, and therefore allowed for a
relatively high water production rate.

The \di{} spacecraft flew by Comet 103P on 4 November 2010.  The
minimum flyby distance was 694~km, occurring at 13:59:47.31 UTC with a
relative speed of 12.3 km~s$^{-1}$ \citep{ahearn11}.  Rather than just
being the fifth comet nucleus to be imaged up close by a spacecraft,
Comet Hartley 2 might also be considered an archetype of the
hyperactive comets.  The flyby images verified the small surface area
of the nucleus \citep[5.24~km$^2$;][]{thomas12-hartley2}, and
\textit{Deep Impact's} IR spectrometer revealed a coma of icy grains
\citep{ahearn11}.  \citet{knight12-thisissue} discovered tail of OH and NH,
evident on $10^4$~km scales, which they conclude are derived from
small grains of ice that were accelerated down the tail before
completely sublimating.  \citet{bonev12} observed enhanced heating of
the water gas 20--150 km down the tail, which they also attribute to
sublimating ice grains in the tail.  The icy coma hypothesis of
\citet{lisse09} seems to be qualitatively valid.  What remains is a
quantitative verification that the ice grains in the coma are a
significant source of water gas.

\citet{ahearn11} also discovered bright point sources distributed
around the nucleus of Hartley 2 in visible wavelength images.  The
fluxes of these sources are consistent with a population of objects
centimeter-sized or larger.  \citet{ahearn11} could not determine if
these large particles were icy or refractory.  In the following paper,
we will attempt to discern the nature of these point sources and if
they might be the cause of Hartley 2's hyperactivity.

The point sources around Hartley 2 are found in visible wavelength
images taken with both the High Resolution Instrument (HRI) and the
Medium Resolution Instrument (MRI).  Their spatial distributions and
densities vary, but they are not background stars; celestial sources
are significantly streaked in images taken while the spacecraft-comet
distance, $\Delta$, was less than $10^4$~km.  That the HRI detected
these point sources is significant.  The HRI visible camera (HRIVIS)
is not in optimal focus, and has an accordingly large and
doughnut-shaped point spread function (PSF) \citep{lindler07}.  Thus,
unresolved sources are easily discriminated from cosmic ray impacts
and hot pixels.  None of the point sources appear to be resolved,
therefore they must be smaller than the FWHM of the HRIVIS deconvolved
PSF (approximately 3~m at $\Delta=700$~km).

In addition to point sources, streaks of various lengths are present
in many of the images.  The lengths of the streaks are correlated with
exposure time (longer exposures produce longer streaks), and inversely
proportional to instrument pixel scale (streaks in HRIVIS images are
longer than in MRI).  The streaks are parallel to each other, and the
orientation is consistent with the spacecraft's motion as it tracks
103P's nucleus.  Stars and other distant objects produce streaks of
uniform length and orientation, but we find a variety of streak
lengths in a single image, down to a few pixels long, which indicates
they are spatially close to the nucleus and not celestial sources.

Altogether, the above observations indicate that the nucleus of Comet
103P is surrounded by a coma of thousands of large particles.  We will
use the term ``particle'' to refer to the observed point sources in
this paper.  \citet{ahearn11} demonstrated that these particles are
centimeter-sized or larger.  Large particles are not unique to this
comet, but these observations are unique to large particles.  These
are the first images of sub-meter particles from a comet seen as
individual objects.  Prior images in visible/infrared light
\citep[e.g., comet dust trails;][]{sykes92, ishiguro02, reach07} and
with radar \citep{harmon04} probed collections of many large
particles.  The only other observations of large individual particles
from comets that we can think of are those of meteors (from meteor
streams associated with particular comets) and milligram particle
impacts on spacecraft.  Although the latter examples may only be of
order 0.1--1~mm in size.

In the following paper, we present our methods for detecting and
measuring the photometric properties of the large particles in the
coma of Comet 103P.  We will convert their fluxes into sizes, discuss
their spatial distribution, attempt to constrain their composition,
and compare them to large particles observed in radar observations of
the comet.  Assuming they are icy, we also constrain their
contribution to the comet's total water production rate.

\section{Observations}
Images of Comet 103p Were taken with \di{}'s Medium Resolution
Instrument and High Resolution Instrument CCD cameras.  Both cameras
have $1024\times1024$ pixel arrays; the HRIVIS pixel scale is
0.413\arcsec~pixel$^{-1}$ (2~\textmu{}rad), and the MRI pixel scale is
2.06\arcsec~pixel$^{-1}$ (10~\textmu{}rad).  The instruments are described
by \citet{hampton05}, and their calibration by \citet{klaasen08,
  klaasen12}.  The MRI and HRIVIS data are available in the NASA
Planetary Data System (PDS) archive \citep{mclaughlin11-dixi-hri,
  mclaughlin11-dixi-mri}.  Images taken with the MRI are labeled with
the prefix mv, and for the HRIVIS the prefix is hv.  We keep this
convention throughout the paper.  We used a pre-PDS data set, but the
only significant difference is the absolute flux calibration constant,
which we account for in our photometry.

Optimal focus for the HRIVIS instrument occurs about 6~mm before the
focal plane.  For this instrument, it is often beneficial to work with
spatially restored (i.e., deconvolved) images.  We follow the method
of \citet{lindler07}, using the HRIVIS PSF from the EPOXI mission
\citep{barry10}, to restore raw HRIVIS images to near
diffraction-limited resolution.  The HRIVIS images are deconvolved
(Fig.~\ref{fig:hri-process}b) with the Richardson-Lucy (R-L) method
modified to handle non-Poisson CCD readout noise \citep{snyder93,
  lindler07}.  Note that R-L restoration methods conserve flux.

The large particles are easily seen in MRI and HRIVIS images at
$\Delta \lesssim 10^4$~km.  Outside of this range, the particles
become increasingly faint, overwhelmed by the diffuse coma's surface
brightness, and confused with stars, which are unstreaked point
sources when the spacecraft-comet distance is large.  In a manual
search, the earliest image in which we have identified particles is
HRIVIS image hv5000096, taken 22,400~km from the nucleus (CA-30~min,
1500.5~ms exposure time).  The first MRI image in which we have
positively identified particles is mv5002025 (8580~km, CA-11~min,
500.5~ms).

Most images at $\Delta<10^4$~km use the CLEAR1 filter.  For sources
with solar-like spectra, the CLEAR1 filters have effective wavelengths
of $\lambda_e = 0.625$~\micron{} for HRIVIS, and 0.610~\micron{} for
MRI, and a FWHM of $\approx0.5$~\micron.  The absolute calibration
uncertainties for the CLEAR1 filters are 5\% for HRIVIS, and 10\% for
MRI \citep{klaasen08}.  Large particles have not yet been identified
in the HRI infrared spectrometer (HRIIR) scans \citep{protopapa11},
but they may be discovered in the future when we can predict the
location of the brightest particles in IR data using their MRI and
HRIVIS derived 3D positions and velocities \citep{hermalyn12}.

A summary of all images used in this paper is presented in
Table~\ref{tab:data}.  Note that the HRIVIS images used in this paper
are only a small sub-set of the HRIVIS particle observations.  We have
limited our investigation to the closest-approach images because of
the great amount of time that must be taken to analyze the HRIVIS
data.  We also do not use any of the 50.5~ms MRI images in the PDS
archive.  These images were taken along with the 500.5~ms images, but
few particles are found in these images due to the decreased
sensitivity and great distances from the nucleus.

\section{Detection, photometry, and completeness}
\subsection{MRI images}\label{sec:mri-phot}
We automatically detect and measure all point sources throughout all
MRI images taken within $\Delta<5200$~km of the comet nucleus using
DAOPHOT as included in the IRAF software package \citep{tody93}.  In
summary, we: (1) estimate the PSF of the instrument and verify that it
accurately measures particle fluxes; (2) remove the diffuse coma from
the images; (3) mask the nucleus and bright jets; (4) additionally
mask stars in images at large $\Delta$; (5) measure the completeness
of our particle census (i.e., detection efficiency); (6) estimate the
photometric uncertainties, and correct for any biases; and, (7) clean
bad PSF fits and crowded regions from our photometry lists.  Below, we
provide details on our methods.

Particles were detected by searching for point sources (FWHM
$=1.5$~pixels) with a peak amplitude greater than $5\sigma$ above the
local background: $1\sigma$ per pixel $\approx1.1 \times10^{-3}$,
$0.52\times 10^{-3}$, and $0.13\times 10^{-3}$
W~m$^{-2}$~\micron$^{-1}$~sr$^{-1}${} for 41, 121, and 501~ms frames,
respectively.  Twenty-five isolated and bright point sources were
selected by hand from image mv5004056 to estimate the MRI PSF.  The
sources were fit by DAOPHOT with a variety of functions, and the
best-fit PSF used a Moffat function \citep[$\beta$ parameter of
  1.5;][]{moffat69} with a look-up table of empirical residuals.  We
compared DAOPHOT-derived fluxes of bright isolated particles to those
measured with aperture photometry.  The values agreed, with a mean
error of $\Delta F_\lambda / F_\lambda = -1.9\%$, where $\Delta
F_\lambda$ is the difference between the DAOPHOT flux and the aperture
flux ($F_{\lambda,\rm DAOPHOT} - F_{\lambda,\rm aper}$).

\begin{figure}
  \ifincludegraphics
  \begin{center}
    \includegraphics[width=\columnwidth]{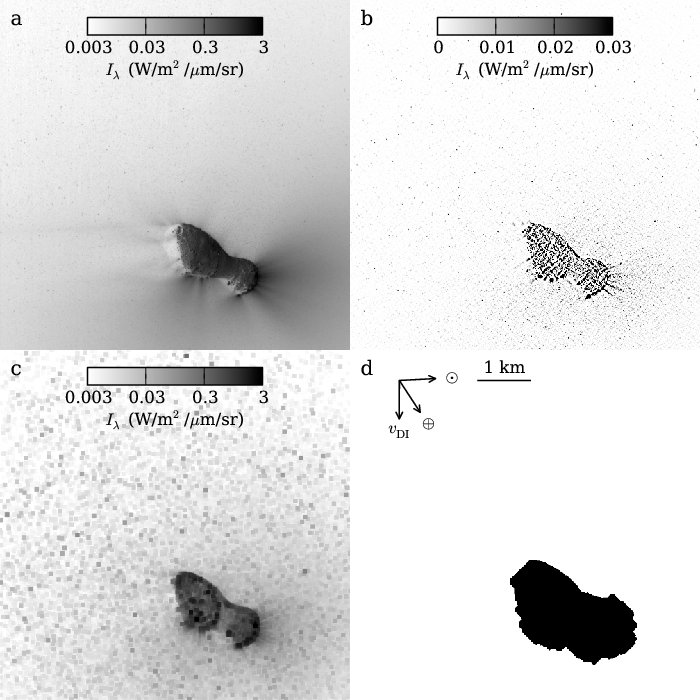}
  \end{center}
  \fi

  \caption[MRI processing]{The process by which we remove the
    background and mask the nucleus in MRI images.  (a) A fully
    calibrated image, mv5004046.  (b) The median-filter subtracted
    image.  (c) Image of the morphological gradient of image b.  (d)
    The final mask, defined where image c $>0.1$
    W~m$^{-2}$~\micron$^{-1}$~sr$^{-1}$.  The mask has been dilated by
    30~pixels to fill gaps and cover some additional coma.  The image
    scale and orientation are given in panel d.  The projected vectors
    are comet-Sun ($\odot$), comet-Earth ($\oplus$), and spacecraft
    velocity ($v_{\rm DI}$). \label{fig:mri-process}}
\end{figure}

DAOPHOT iteratively estimates the background as it fits each point
source and groups of point sources, but we found---especially for
detections closer to the nucleus---that subtracting an initial
estimate of the background improved DAOPHOT's ability to find and
measure sources.  Furthermore, the MRI CCD has an instrumental
background that varies row-by-row at the 1--4~DN level (the MRI and
HRIVIS calibration constants for CLEAR1 are $3.527\times 10^{-5}$ and
$1.209\times 10^{-4}$ W~m$^{-2}$~\micron$^{-1}$~sr$^{-1}$~DN$^{-1}$~s,
respectively).  The EPOXI pipeline includes a routine to detect and
remove this background, but it is not executed on images dominated by
coma, such as those in our analysis.  We derive our background
estimate for each image using two median filters (i.e., low-pass
filters).  We first apply a $3\times25$~pixel (row $\times$ column)
filter that defines the background stripes, as well as much of the
coma.  A second $25\times3$~pixel filter defines most of the residual
coma, especially where it is extended along image columns.  A
median-filter subtracted image is presented in
Fig.~\ref{fig:mri-process}b.

Bright jets and the nucleus complicate the PSF fitting.  Therefore, we
mask these regions on the image.  The mask is derived from a
morphological gradient with an $11\times11$~pixel uniform structuring
element, which effectively generates an image of peak-to-peak values
in a moving $11\times11$~pixel box (Fig.~\ref{fig:mri-process}c).  We
threshold the gradient image at a value of 0.1
W~m$^{-2}$~\micron$^{-1}$~sr$^{-1}$, then dilate the mask by 30~pixels
to fill small gaps between features, and cover some of the nearby jets
(Fig.~\ref{fig:mri-process}d).

DAOPHOT will occasionally fit streaks in the MRI images with PSFs.
Most of these fits have poor residuals, and are removed before the
final analysis (described below).  But we found several streaks
persisting into the final analysis in images taken at large distances
from the comet ($\Delta\gtrsim5000$~km).  These streaks were stars and
not particles.  DAOPHOT seemed more likely to fit streaks in these
images because the stars are much brighter than the particles, and are
only streaked several pixels or less.  We generated an additional mask
by searching for elongated sets of pixels above the background in
order to remove streaked celestial objects from these images before
processing with DAOPHOT.  At closer comet-spacecraft distances, stars
are streaked over a greater number of pixels (up to a few hundred
pixels at closest approach), and are easily rejected without the mask.

\begin{figure}
  \ifincludegraphics
  \begin{center}
    \includegraphics[width=1.0\columnwidth]{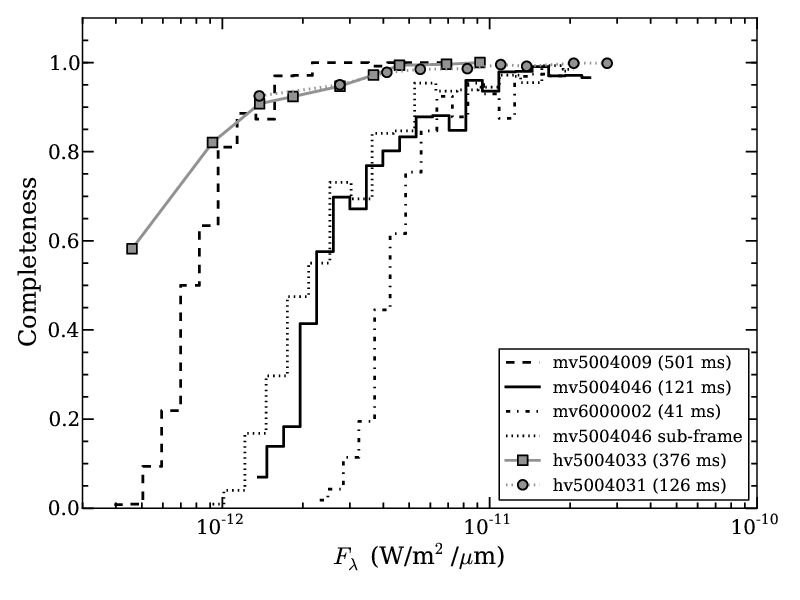}
  \end{center}
  \fi

  \caption[Completeness]{Example point source detection completeness
    functions.  All images were taken within $\Delta=694-704$~km,
    except mv5004009 (501~ms) taken at $\Delta=1858$~km.  The
    mv5004046 sub-frame is chosen to match the same area as covered in
    hv5004031.  The HRIVIS test particles are selected from one of 9
    specific fluxes (marked by squares and circles), whereas the MRI
    test particles are selected at random from a range of fluxes.  The
    lowest of the HRIVIS bins is
    10,000~e$^-$. \label{fig:completeness}}
\end{figure}

We verified the completeness (i.e., detection efficiency) of our
method by inserting artificial point sources into each image,
re-executing our scripts on the new images, and examining the output
to determine if the artificial sources were detected.  The
completeness is then the fraction of detected artificial sources per
flux bin.  We inserted 1\% of the number of detected point sources
(with a minimum of 10 new particles) uniformly over the image but
avoiding the nucleus.  The fluxes were picked from a distribution
uniform in log-space, based on the minimum and maximum fluxes measured
in the image.  The process is repeated, each time starting with the
original image, until 3000 artificial particles have been tested.
Examples of our MRI completeness test results are shown in
Fig.~\ref{fig:completeness}.  The completeness is directly correlated
with exposure time; in general, it is easier to find and measure
fainter particles in images with longer exposures.  In our MRI
population studies, we will only consider particles brighter than the
80\% completeness level for each image (listed in online
Table~\ref{tab:fits} as $F_{\lambda, {\rm min}}$).

\begin{figure}
  \ifincludegraphics
  \begin{center}
    \includegraphics[width=1.0\columnwidth]{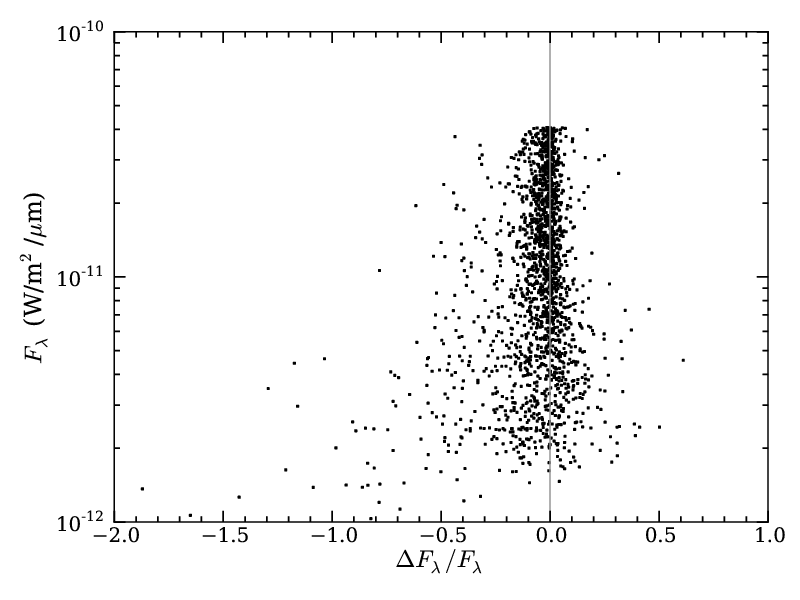}
  \end{center}
  \fi

  \caption[MRI photometry error]{Input flux, $F_\lambda$, plotted
    versus DAOPHOT error, $\Delta F_\lambda / F_\lambda$, for all
    artificial particles inserted in image mv5004046.  Most fluxes are
    measured to within 8\% ($1\sigma$) of their true value, with a
    slight skew toward negative errors. \label{fig:mri-photerr}}
\end{figure}

The completeness tests can be used to derive the photometric
uncertainties.  As an example, we plot the input flux versus the
DAOPHOT flux error for all artificial particles added to image
mv5004046 (Fig~\ref{fig:mri-photerr}).  In this image, the mean flux
error $\overline{\Delta F_\lambda / F_\lambda} = -8\%$, and the mean
error for all images ranges from $-10$\% to $-2$\%.  However, the
distribution is not Gaussian as there is a significant tail at large
negative errors.  If we instead take the resistant mean (clipping at
$3\sigma$), we find the mean $\Delta F_\lambda / F_\lambda$ ranges
from $-3.5$\% to $-0.7$\%, with per image standard deviations ranging
from 6\% to 10\%.  As discussed above, we also found a slight negative
error when we compared the DAOPHOT photometry to aperture photometry
of isolated bright sources.  To account for this effect, we will
increase the fluxes of all particles by 2\%.  In addition, we adopt
8\% as our particle flux uncertainty.

\begin{figure}
  \ifincludegraphics
  \begin{center}
    \includegraphics[width=\columnwidth]{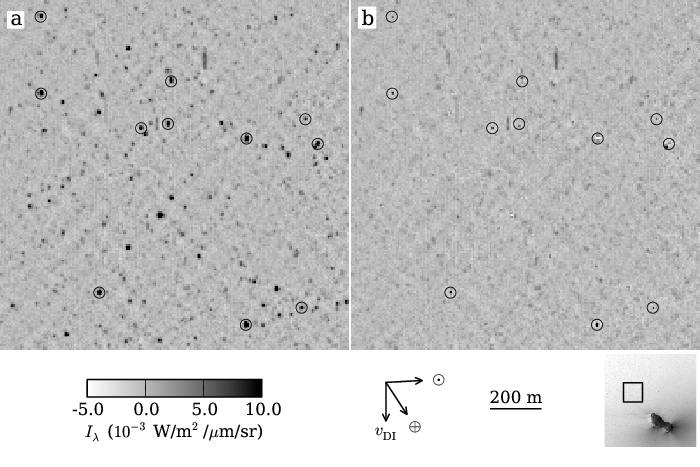}
  \end{center}
  \fi

  \caption[MRI point source removed 1]{(a) A detail of image
    mv5004046, after background coma subtraction.  (b) The same image
    after being cleaned of point sources detected and fit by DAOPHOT.
    In both images, particles with large residuals or low
    signal-to-noise flux ratios have been circled.  Similar to
    Figure~\ref{fig:mri-process}, the image scale and orientation are
    shown.  The square in the lower right image shows the field of
    view of panels a and b with respect to the nucleus.
  \label{fig:mri-pscleaned-1}}
\end{figure}

\begin{figure}
  \ifincludegraphics
  \begin{center}
    \includegraphics[width=\columnwidth]{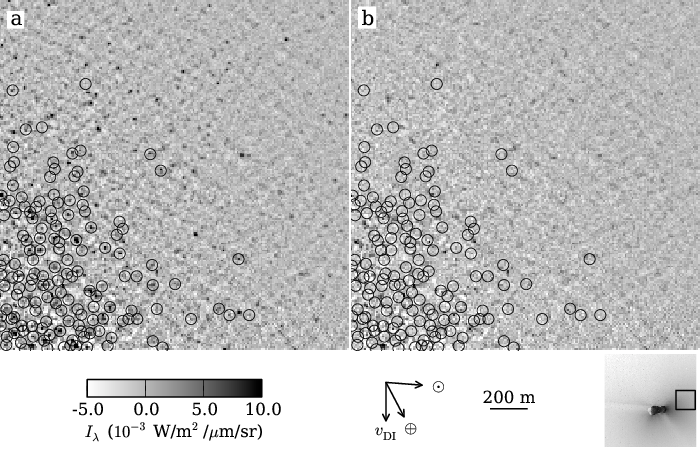}
  \end{center}
  \fi

  \caption[MRI point source removed 2]{The same as
    Fig.~\ref{fig:mri-pscleaned-1}, but for a region of high point
    source density in image mv5004029.  \label{fig:mri-pscleaned-2}}
\end{figure}

The large flux errors ($|\Delta F/F| > 10\%$) are likely misidentified
artificial particles, artificial particles that have merged with
brighter particles in the PSF fitting process, and PSF over/under
fitting.  Any particles with significantly large flux errors must be
identified and removed from our photometry lists so that they do not
affect the final results.  We clean the DAOPHOT output by rejecting:
(1) bad PSF fits as identified by DAOPHOT (i.e., reduced $\chi^2 >
2.0$); and (2) any particles with a total residual in a
$3\times3$~pixel box greater than $3\sigma$ from the background.  For
the 41~ms images $62-78\%$ and $0-2\%$ of the detected sources were
rejected by the two criteria, respectively.  For the 121~ms images the
rejection rates were $5-32\%$ and $2-5\%$; for 501~ms, the rates were
$0-1$\%, and $3-20$\%.  Occasionally DAOPHOT fits portions of the
particle streaks, and these criteria help remove such spurious fits
from the photometry lists.  Examples of point source cleaned images
are shown in Figs.~\ref{fig:mri-pscleaned-1} and
\ref{fig:mri-pscleaned-2}.  After point source cleaning, sources not
fit by DAOPHOT are evident.  These particles can be accounted for in a
statistical sense with the image's completeness function, e.g., when
we are displaying the flux distribution of the particles, we divide
the measured flux distribution by the completeness curve to show the
true flux distribution.

In Fig~\ref{fig:mri-pscleaned-2} we specifically show a region of high
point source density, and the large number of bad-PSF fits resulting
from point source crowding near the nucleus ($>0.025$ pixel$^{-1}$).
These sources are primarily rejected by the $\chi^2$ criterion.
Regions with a rejection rate $>20\%$ in a $25\times25$~pixel area
will be masked from our analyses.

\subsection{HRIVIS images}
\begin{figure}
  \ifincludegraphics
  \begin{center}
    \includegraphics[width=0.9\columnwidth]{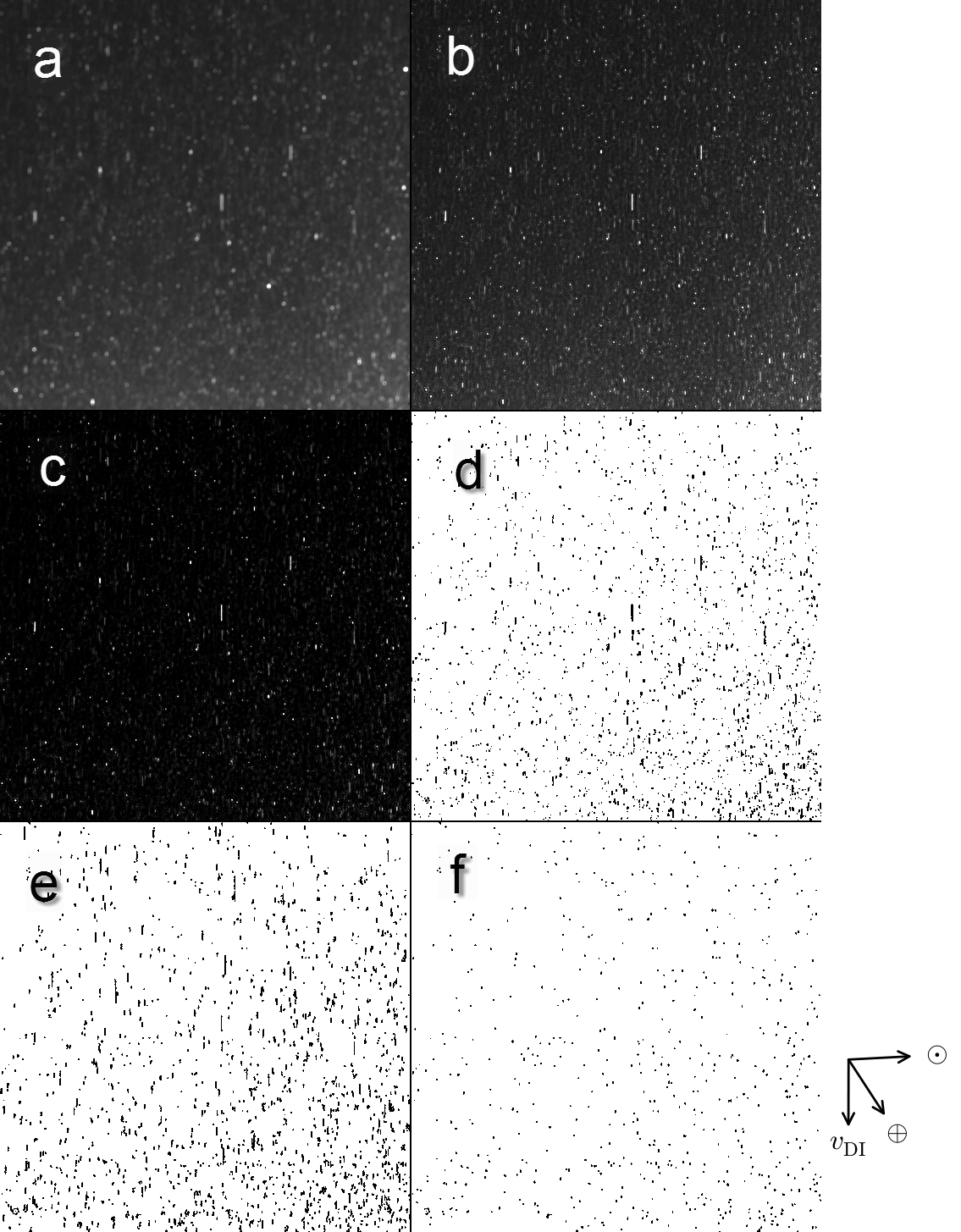}
  \end{center}
  \fi

  \caption[HRI processing]{(a) Original HRIVIS image hv5004030. (b)
    Image deconvolved with the R-L method. (c) Image b after
    background subtraction.  (d) Mask of pixels greater than
    1000~e$^-$ above the background. (e) Mask of detected sources.
    (f) Mask of the sources used to compute the flux distribution
    slope (here, length 5 pixels or smaller and total flux greater
    than 20,000~e$^-$). \label{fig:hri-process}}
\end{figure}

The analysis of the HRIVIS images required an approach independent
from the MRI analysis.  Figure~\ref{fig:hri-process}a shows an HRIVIS
image of the large particles and Fig.~\ref{fig:hri-process}b is the
restored (i.e., deconvolved) image.  The broad defocused PSF and the
longer trails resulting from parallax motion make PSF fitting even
more difficult than in the MRI images.  Our approach is to instead
detect and measure a particle's brightness in restored images.  In
summary, we: (1) iteratively remove the background from each image; (2)
detect point sources and streaks in the restored images; (3) measure
particle fluxes; and, (4) measure the completeness of our particle
detection scheme.  Below we detail our methods.

We use an iterative technique to remove the background before final
object detection and photometry.  The initial background is set to the
deconvolved image filtered with a $51\times51$~pixel median filter.
We then fit a smooth surface (using cubic splines) to the filtered
image to obtain our background image.  The objects within the image
will bias the median filter and the resulting image overestimates the
background.  To minimize this bias, we repeat the process after
ignoring all pixels more than 400 e$^-$ above the background (object
pixels) during median filtering \citep[1 e$^-$~s$^{-1}$ =
  $4.32\times10^{-6}$ W~m$^{-2}$~\micron$^{-1}$~sr;][]{klaasen12}.  We
repeat this process three times to obtain and remove our final
background (Fig.~\ref{fig:hri-process}c).

We detect the objects in the image by creating a mask of all pixels
more than 1,000 e$^-$ above the background level
(Fig.~\ref{fig:hri-process}d).  We selected the 1,000 e$^-$ threshold
by visual inspection of the restored image to avoid detection of
restoration artifacts that result from ``ringing'' around the brighter
objects \citep{lindler12}.  We define contiguous masked pixels as part
of the same object.  Visual inspection of this mask shows many
instances where a trailed particle near the detection threshold is
split into multiple objects because of noise and potential variations
in particle brightness during the time of the exposure.  To avoid this
problem, we examine each detected object (starting with the brightest)
and adding all pixels within a factor of 3 of the brightest pixel in
the object.  We define the resulting contiguous dark regions as our
detected objects (Fig.~\ref{fig:hri-process}e).

We perform photometry on the detected particles by starting with the
particle mask, and adding the flux from their nearest-neighbors (i.e.,
we grow the mask by 1 pixel, and sum the pixels together).
Non-linearities are a concern when restoring an image with non-linear
deconvolution algorithms \citep{lindler94}.  Fainter point sources
will be broader than brighter point sources in a restored image.  We
can minimize any non-linearities by increasing the size of the region
used for photometry.  However, this results in more contamination of
our results by nearby sources. To test for non-linearity we increased
the size of the photometry regions by adding additional neighbors
(i.e., including next-nearest neighbors, etc.).  These larger regions
showed that non-linearity artifacts are absent from particles brighter
than 10,000~e$^-$ ($F_\lambda = 1.4\times10^{-12}$ and
$4.6\times10^{-13}$ W~m$^{-2}$~\micron$^{-1}${} for exposure times of
126 and 376~ms).  On average, the effective aperture for point sources
is approximately a 3~pixel radius circle.

As done with the MRI analysis, we measure the completeness by
inserting artificial objects in the images of varying brightness and
parallax length.  These artificial objects are added to the raw images
before deconvolution.  The completeness test also gives additional
evidence that our measured fluxes are linear with object brightness in
the range of brightness being considered.  For the longer trails, the
completeness decreased rapidly with flux.  We therefore decided to
ignore particles with a length more than 5 pixels in the restored
image.  In effect, this criterion limits the line-of-sight distance of
the particles from the nucleus to $<1-2$~km at closest approach
($\Delta=694$~km), and $<2-3$~km at $\Delta=900$~km.  If the flux
distribution is independent with distance from the nucleus, then the
relative distribution of number versus flux will remain unchanged.
Limiting the length of the objects also has added benefits.  Long
streaks are much more likely to be a combination of multiple particles
and their photometry is also more strongly affected by errors in the
background estimation (they cover more pixels).  Examples of our
HRIVIS completeness test results are presented in
Figure~\ref{fig:completeness}.  There is little difference between the
126~ms and 376~ms completeness curves in the example.  Therefore our
completeness is not as strongly affected by background noise as it is
by other factors, e.g., point source crowding and deconvolution
artifacts.  In addition, our HRIVIS photometry is generally limited by
the linearity criterion of $F>10,000$~e$^-$.

\section{Particle fluxes}
\subsection{Flux distribution}\label{sec:flux-dist}
\begin{figure}
  \ifincludegraphics
  \begin{center}
    \includegraphics[width=0.85\columnwidth]{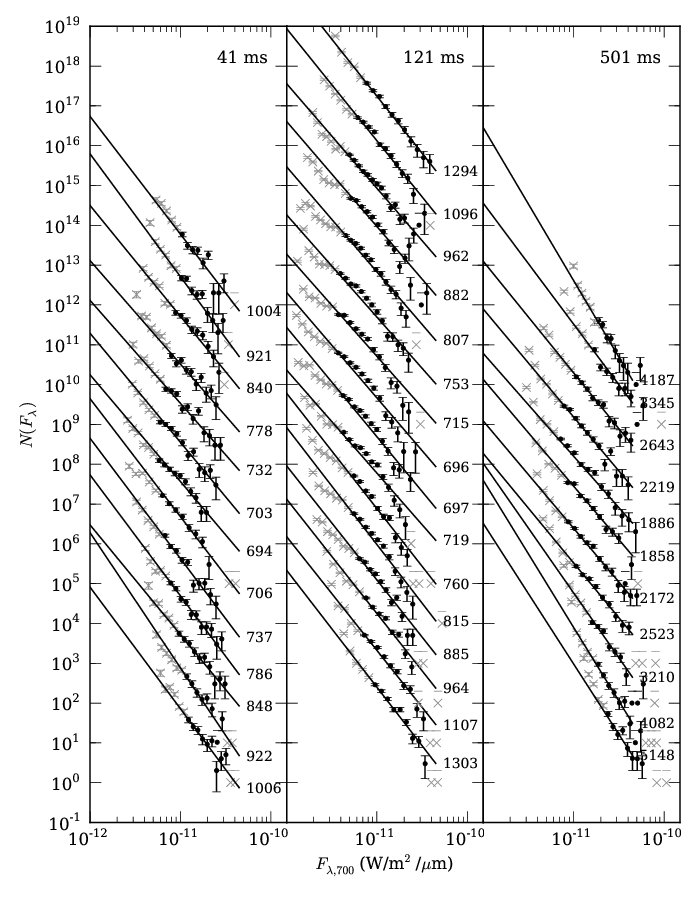}
  \end{center}
  \fi

  \caption[MRI power-law fits]{Flux distributions ($dn/d\log{F}$) and
    their best-fit power-law functions for all full-frame MRI data
    sets listed in Table~\ref{tab:fits}.  The flux distributions have
    been corrected for completeness, and their error bars are based on
    Poisson statistics.  The flux distributions are grouped by their
    integration times, and sorted by time with the pre-closest
    approach images at the bottom.  Each $i$-th flux distribution
    ($i=0, 1, \ldots$) has been offset along the y-axis by $10^i$, and
    labeled with $\Delta$ for clarity.  Additionally, the fluxes have
    been scaled by $(\Delta / 700)^2$ so that they are directly
    comparable to each other.  The gray $\times$ symbols mark
    particles not used in the fit. \label{fig:mri-fits}}
\end{figure}

\begin{figure}
  \ifincludegraphics
  \begin{center}
    \includegraphics[width=0.8\columnwidth]{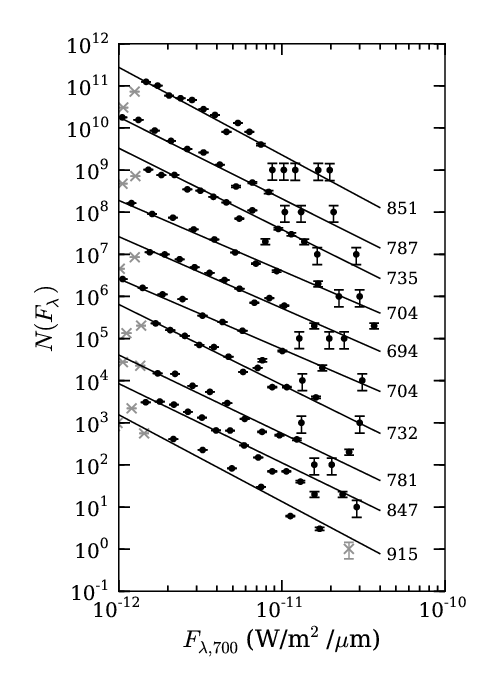}
  \end{center}
  \fi

  \caption[HRI power-law fits]{Same as Fig.~\ref{fig:mri-fits}, but
    for all HRIVIS images listed in Table~\ref{tab:fits}.
  \label{fig:hri-fits}}
\end{figure}

We measure the fluxes of all particles and compute their flux
distributions, $dn/dF$.  The particle fluxes approximately follow a
power-law distribution.  Therefore, we fit each flux distribution with
a function of the form $dn/dF\propto F^\alpha$ using the method by
\citet{clauset09}, which is based on Kolmogorov-Smirnov tests and
maximum likelihood fitting.  We restricted the flux range to particles
brighter than our 80\% completeness estimate.  This restriction
increases the statistical errors in the fits but reduces possible bias
errors resulting from photometry contamination from neighboring
particles and residual background.  Moreover, the fitting method does
not include a correction for completeness, so restricting the fluxes
helps mitigate the effect of incomplete counting; the maximum
completeness correction over our fit range is 20\%, which is
insignificant in comparison to the strong power-law slopes \citep[see
  Fig.~S6 of][]{ahearn11}.  Our results are listed in
Table~\ref{tab:fits} and summarized in Table~\ref{tab:fit-summary}.
In Figs.~\ref{fig:mri-fits} and \ref{fig:hri-fits}, we plot the final
flux distributions and their best-fit trends.  The fluxes are plotted
as a function of $\fnorm\equiv F_\lambda \Delta^2 / 700^2$ so that
they may be directly compared to each other (variation of the phase
angle is small, ranging from 79 to 92\degr, and is therefore not
considered).  The error-weighted mean slopes for the MRI and HRIVIS
frames are $-3.74\pm0.02$ and $-2.85\pm0.03$, respectively.

The MRI and HRIVIS slopes are significantly different from each other.
To verify this difference, we also computed the flux distributions
from MRI sub-frames of images chosen close in time to the HRIVIS
images.  The relative position of the HRIVIS and MRI fields are
constant with time, therefore measuring the flux distribution in the
same field of view is straightforward.  The comparison, however, is
not exact because the frame-to-frame parallax of individual particles
can be significant \citep{hermalyn12}.  After bad-PSF rejection and
completeness tests only $\approx130$ particles could be used to
determine the flux distributions in the MRI sub-frames.  This is a
factor of 3--10 fewer than the number of particles in the HRIVIS
images.  The difference primarily relies on the larger aperture of the
HRI primary, which allows for a fainter point source detection limit
due to the finer resolution and greater light gathering area.  Our
best-fit parameters for the MRI sub-frames are listed in
Table~\ref{tab:fits} and summarized in Table~\ref{tab:fit-summary}.
The uncertainties in the power-law slopes are quite large, ranging
from 0.18 to 0.31, owing to the paucity of particles available in the
MRI sub-frame images.  The weighted mean slope is $-3.59\pm0.07$,
which is shallower than the mean MRI full-frame slope, but they agree
at the $2\sigma$ level.

Taken together, the mean MRI full-frame, MRI sub-frame, and HRIVIS
best-fit power-law slopes show evidence that the flux distribution
steepens with increasing flux.  Inspection of the MRI full-frame
distributions in Fig.~\ref{fig:mri-fits} and the Kolmogorov-Smirnov
probability ($P_{KS}$) values reveals that a power-law size
distribution is not a good fit to most of the flux distributions.
This poor fit is especially true for the 121~ms frames at $\Delta <
900$~km where there is a significant curvature in the log-flux
distributions of these images.  The curvature is non-existent or not
prevalent in the shorter exposure MRI data, the distant MRI data
($\Delta>1000$~km), and the HRIVIS data.

\begin{figure}
  \ifincludegraphics
  \begin{center}
    \includegraphics[width=1.0\columnwidth]{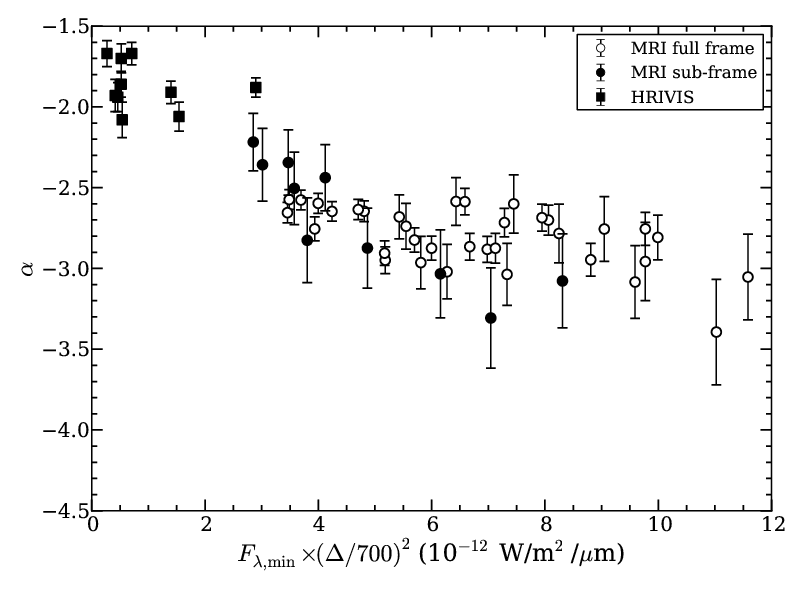}
  \end{center}
  \fi

  \caption[Slope vs. Fmin]{Best-fit power-law slope versus minimum
    flux used in the fit for all images listed in
    Table~\ref{tab:fits}.  \label{fig:slope-vs-fmin}}
\end{figure}

It is important to recognize two points.  First, the HRIVIS
distributions are poorly populated at $\fnorm > 10^{-11}$
W~m$^{-2}$~\micron$^{-1}$, whereas the 121~ms MRI distributions are
well defined up to $\fnorm\approx3\times10^{-11}$
W~m$^{-2}$~\micron$^{-1}$.  This difference is simply a matter of the
larger solid angle observed by the MRI camera as compared to the
HRIVIS camera; a larger solid angle allows for more of the brightest
particles to be detected.  Second, the HRIVIS distributions are valid
to lower fluxes than the MRI distributions (column $F_{\lambda, {\rm
    min}}$ in Table~\ref{tab:fit-summary}), because the larger primary
of the HRI allows for fainter particles to be accurately measured.
Taken together, these differences demonstrate that the two instruments
are measuring two different particle flux regimes.  In
Fig.~\ref{fig:slope-vs-fmin}, we plot our best-fit power-law slopes
versus the minimum flux used in each fit.  A trend is evident; steeper
slopes are correlated with larger \fmin{}, but due to the different
techniques, fit uncertainties, and fields of view involved, the
correlation is not conclusive.  We remind the reader that the
correlation is not due to incomplete photometry because we restricted
our data sets to fluxes where the completeness is better than 80\%.
Given this restriction, none of our completeness corrections are
strong enough to have a significant effect on the best-fit slopes.

\begin{figure}
  \ifincludegraphics
  \begin{center}
    \includegraphics[width=0.8\columnwidth]{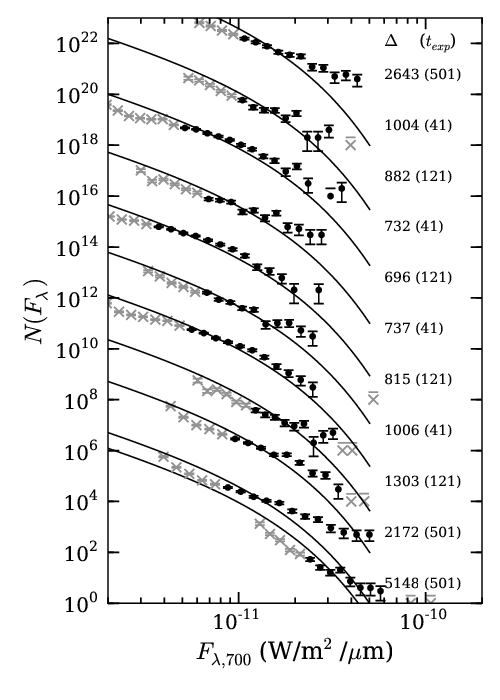}
  \end{center}
  \fi

  \caption[Power-law with exp cut off]{Best-fit power-law with
    exponential cut off, derived from 121~ms MRI images taken within
    $\Delta=1000$~km, compared to selected flux distributions, labeled
    with ``$\Delta$ (exposure time)'' in km and ms.  The histograms
    have been offset by a factor of $10^i$ for
    clarity.  \label{fig:mri-plco-mean}}
\end{figure}

In the absence of any additional information on the flux distribution
we will proceed assuming that the differences between the MRI and
HRIVIS flux distributions reflect the true flux distribution of the
entire particle population.  As alternatives to a power-law
distribution, we also considered a power-law with an exponential cut
off, and a double power-law.  The functional forms are
\begin{linenomath}
  \begin{align}
    \frac{dn}{dF} & \propto F^{\beta_1} \exp{(-F / \gamma_1)}\mbox{ ,}
    \label{eq:plco} \\
    \frac{dn}{dF} & \propto \frac{F}{\gamma_2}^{\beta_2} \left(1 +
    \frac{F}{\gamma_2}\right)^{\beta_3 - \beta_2}\mbox{ ,}
    \label{eq:pl2}
  \end{align}
\end{linenomath}
where $\beta_i$ are power-law slopes, and $\gamma_1$ is the flux at
which the exponential decay reaches a factor of $\exp{(-1)}=0.37$ ,
and $\gamma_2$ is the turn-over flux from the low- to the high-flux
slopes (both $\gamma$ parameters are specified for $\Delta=700$~km).
The double power-law fits did not converge on a single solution.  The
power-law with exponential cut-off function, however, is a better fit
to the closest-approach data ($\Delta<1000$~km), with a mean slope and
cut-off equal to $-3.18\pm0.07$ and
$(5.7\pm0.3)\times10^{-12}$~W~m$^{-2}$~\micron$^{-1}$.  Unfortunately,
these best-fit values are poor fits at larger distances where the flux
distribution better agrees with a power-law
(Fig.~\ref{fig:mri-plco-mean}).  We suggest that the 121~ms
distributions have a systematic effect that causes under-counting of
the largest flux bins.  Again, we note that our completeness
corrections are not strong enough to significantly affect our fits,
and that the distributions as shown in Figs.~\ref{fig:mri-fits} and
\ref{fig:mri-plco-mean} have been corrected for completeness, yet the
curvature remains in the closest approach data.

To test the hypothesis that the power-law slope is affected by the
source density, we generated completely synthetic data sets with
DAOPHOT, based on our MRI PSF.  We created 8 images each with the same
power-law flux distribution ($dn/dF\propto F^{-3.5}$) and maximum
point source flux ($\fmax=1\times10^{-10}$ W~m$^{-2}$~\micron$^{-1}$),
but \fmin{} varied from $5\times 10^{-12}$ down to $1.5\times
10^{-13}$ W~m$^{-2}$~\micron$^{-1}$.  We inserted the same number of
point sources per logarithmic flux bin, but because each image's flux
distribution spanned a different range, the total point source density
varied from $4\times 10^{-4}$~pixel$^{-1}$ up to 2.6~pixel$^{-1}$.  We
then measured each image's flux distribution with DAOPHOT in a manner
similar to our MRI images.  Once the input column density reached
0.08~pixel$^{-1}$, our best-fit slopes steepened from the input $-3.5$
down to a minimum of $-4.7$ at the highest densities.  Thus, the
rollover in the flux distribution seen in the 121~ms MRI images at
$\Delta < 1200$~km could be a manifestation of this effect.

\begin{figure}
  \ifincludegraphics
  \begin{center}
    \includegraphics[width=1.0\columnwidth]{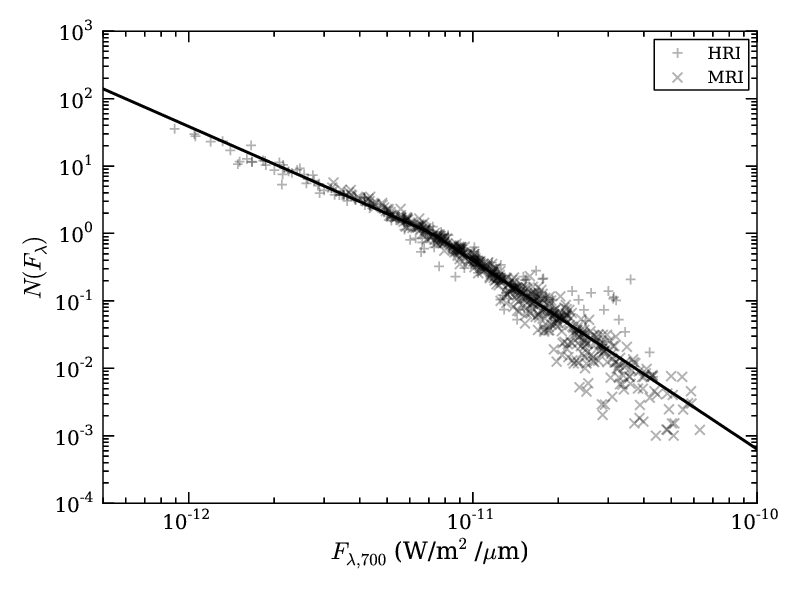}
  \end{center}
  \fi

  \caption[Master flux distribution]{Combined HRIVIS and MRI flux
    distributions ($F_\lambda\geq\fmin$), scaled to
    $N(F_\lambda=7.2\times10^{-12}) = 1$.  For the MRI data, only
    full-frame distributions from $\Delta>900$~km are shown.  The
    solid line is our best-fit broken power-law
    (Eq.~\ref{eq:broken}). \label{fig:master-fluxdist}}
\end{figure}

The slopes derived from the distant MRI and the 41~ms MRI images are
consistent: $\alpha=-3.80\pm0.02$ and $-3.83\pm0.05$, respectively.
They agree despite the wide range in the number of particles fit
(100--1000) and spacecraft-comet distances (900--5000~km, although the
best constraints are within 3000~km).  Therefore, we consider these
results to be robust and representative of the true flux distribution
for $\fnorm\gtrsim1\times10^{-11}$ W~m$^{-2}$~\micron$^{-1}$.  The
shallower HRIVIS slopes appear valid for $1\times10^{-12} \lesssim
F_\lambda \lesssim 1\times10^{-11}$.  We see no reason to trust one
data set over the other, or to assume that a single power-law would be
valid for all fluxes measured.  Therefore, we will use a broken
power-law for the remainder of the paper:
\begin{equation}
  \frac{dn}{dF} \propto F^\alpha
  \begin{cases}
    \alpha = -2.85 & \text{ for } \fnorm < 7.2\times10^{-12} \text{ W~m$^{-2}$~\micron$^{-1}$} \\
    \alpha = -3.80 & \text{ for } \fnorm \geq 7.2\times10^{-12} \text{ W~m$^{-2}$~\micron$^{-1}$} \\
  \end{cases},\label{eq:broken}
\end{equation}
where $\fnorm=7.2\times 10^{-12}$ is the break in the power-law,
$-2.85\pm0.02$ is the average HRIVIS slope, and $-3.80\pm0.02$ is the
average MRI slope, measured from images at $\Delta>900$~km to avoid
the apparent crowding effects at closest approach.  The break was
derived by least-squares fitting the combined MRI and HRIVIS data
sets.  The broken power-law and the combined HRIVIS and MRI flux
distributions are presented in Fig.~\ref{fig:master-fluxdist}.  This
model is a good match to the ensemble (HRIVIS + MRI) data set.

\subsection{Total particle flux}\label{sec:total-flux}
\begin{figure}
  \ifincludegraphics
  \begin{center}
    \includegraphics[width=1.0\columnwidth]{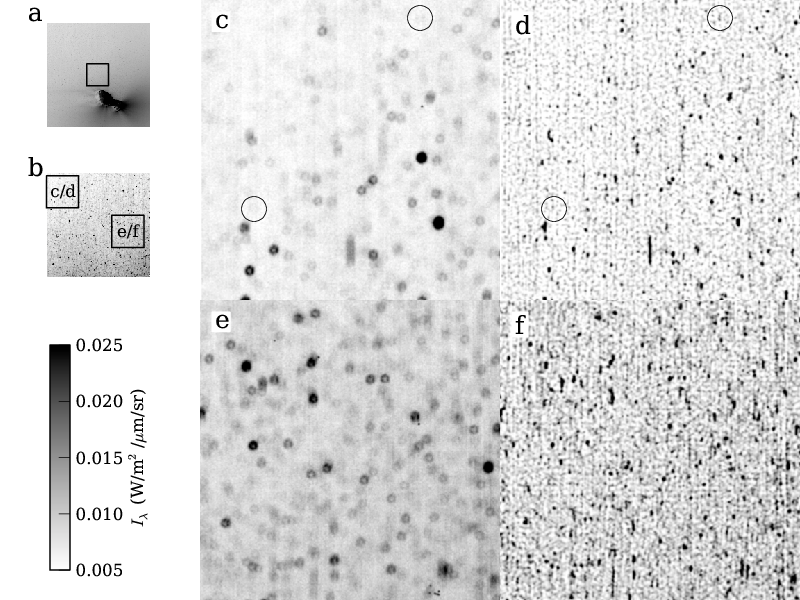}
  \end{center}
  \fi

  \caption[HRI detail]{Two sub-frames of HRIVIS image hv5004031
    ($\Delta=694$~km) showing the crowded field of particles: a) an
    MRI context image (mv5004046) with the approximate HRIVIS field of
    view outlined with a box (image is log scaled from 0.001 to 1.0
    W~m$^{-2}$~\micron$^{-1}$~sr$^{-1}$); b) HRIVIS image, sub-frames
    c/d and e/f are outlined and labeled; c) HRIVIS sub-frame; d) the
    same as c, but deconvolved to enhance the spatial resolution; e)
    another HRIVIS sub-frame; f) panel e, deconvolved.  Two particles
    with fluxes near $1-2\times 10^{-13}$ W~m$^{-2}$~\micron$^{-1}${}
    have been circled.  The HRIVIS field of view is 1.4~km, and the
    sub-frames are 0.42~km.  \label{fig:hri-detail}}
\end{figure}

Due to the great numbers of large particles, we can only accurately
measure the total scattered flux from the brightest particles in each
image.  Our best estimate of the fraction of the coma flux
attributable to large particles is 0.11\% for fluxes ranging
$0.9-4.7\times10^{-11}$~W~cm$^{-2}$~\micron$^{-1}$ (row labeled ``MRI
(distant)'' in Table~\ref{tab:fit-summary}).  This estimate was
derived from images with relatively little point source crowding
($\Delta>2000$~km), and a moderate number of particle measurements
($N>300$ particles).  In order to build a complete census of the
particles, we need to extrapolate the results from that limited range
down to the faint end of the distribution.  We find that the brightest
particles observed throughout the closest approach images are
consistently near $\fnorm =
4.5\times10^{-11}$~W~cm$^{-2}$~\micron$^{-1}$.  On the other end, the
faintest particles we can manually find and measure have fluxes of
order $\fnorm\sim10^{-13}$~W~cm$^{-2}$~\micron$^{-1}$
(Fig.~\ref{fig:hri-detail}).  If the very steep flux distribution we
derived in \S\ref{sec:flux-dist} extends from $\fnorm=1\times10^{-12}$
down to $1\times10^{-13}$~W~cm$^{-2}$~\micron$^{-1}$, there should be
several million particles per image.  However, at such great pixel
densities (1 particle per 17 pixel area) it should be very difficult
to find faint isolated particles (the core of the HRIVIS native PSF
has an area of 113 pixels).  Furthermore, if we let the flux
distribution continue down to
$1\times10^{-14}$~W~cm$^{-2}$~\micron$^{-1}$, the large particles
account for 100\% of the coma flux, leaving no room for any fainter
particles (millimeter sized and smaller).  Therefore, the flux
distribution must change or be truncated at fluxes fainter than
$\fnorm=1\times10^{-12}$~W~cm$^{-2}$~\micron$^{-1}$.  The uncertainty
in the lower flux limit is a major source of error in all estimates of
the total number, flux, cross section, and similar derived quantities.
We list the results from our extrapolations in
Table~\ref{tab:extrapolate}.  We estimate that the large particles
account for 2--14\% of the total coma flux near the nucleus.

\begin{figure}
  \ifincludegraphics
  \begin{center}
    \includegraphics[width=1.0\columnwidth]{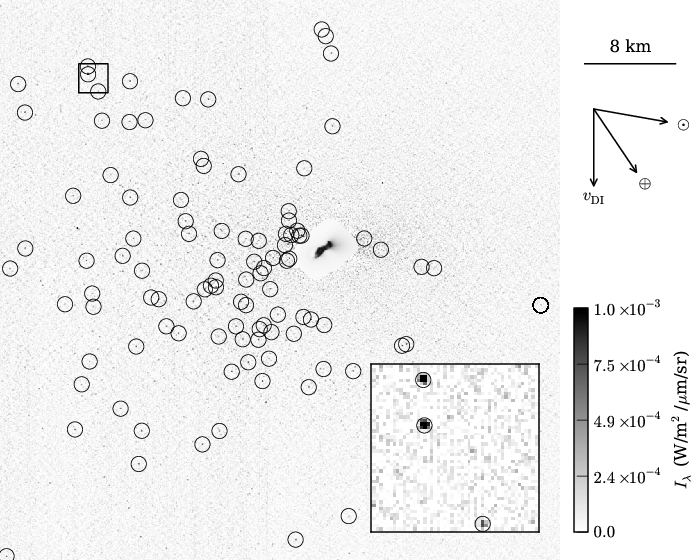}
  \end{center}
  \fi

  \caption[Particle coma extent]{Image mv5002051, background coma
    removed.  Point sources identified by DAOPHOT with fluxes greater
    larger than the image's 80\% completeness limit are circled to
    show the extent of the large particle coma (particles below this
    limit can still be seen by eye).  An image of the nucleus has been
    inserted into the central masked region.  The box marks the
    location of the inset image.  Stars, most of which have been
    automatically masked, are streaked over 13~pixels in this
    image.  \label{fig:mri-extent}}
\end{figure}

If we take the the first image in which particles are easily seen,
mv5002051 at $\Delta=5148$~km (Fig.~\ref{fig:mri-extent}), we can
estimate the total number of particles within a 20.6~km radius (the
largest circular aperture centered on the nucleus that fits within the
image).  We find a total coma flux of $(2.15\pm0.22)\times10^{-7}$
W~m$^{-2}$~\micron$^{-1}$ (includes the bright jets, but not the
nucleus) and assume that 2--14\% of that flux is attributable to large
particles.  The results are presented in Table~\ref{tab:extrapolate}
and will be used below when we estimate the cross section, mass, and
water production rate of the large particles.

Up to now, we have not addressed streaked particles in our MRI images.
If a significant fraction of the flux in our images at
$\Delta\approx2400$~km is contained within streaked particles then our
estimates on the total flux and number of large particles will be
systematically low.  However, we find that few particles are streaked
and the correction to include any possible streaks is negligible.  To
demonstrate, we developed a Monte Carlo simulation that uses the
spacecraft position and velocity to estimate the volume of space that
is smeared.  Within a 20.6~km projected radius, 2\% the field-of-view
is smeared over $\geq1.5$ MRI pixels, and only particles outside of
120~km are smeared this much.  If the particle density follows a
$r^{-2}$ profile, the fraction of smeared particles reduces to
effectively zero.

\section{Particle size and composition}\label{sec:sizes}
In the absence of any compositional information on these large
particles, we assume two cases to demonstrate their likely range of
sizes.  First, we will consider that the particles are refractory, and
photometrically behave like comet nuclei (the ``dusty'' case).  Then,
we will consider that the particles are icy, and behave like the icy
satellite Europa (the ``icy'' case).  We stress that these two cases
are examples only.  They may not reflect the true nature of the
particles, but they do yield useful limits on the particle sizes.  We
will show that the particles are much larger than the
0.1--100~\micron{} sizes typically considered in comet dust.  We
purposefully avoid interpreting the large particles with phase
functions that have been derived for comet comae.  Light scattered by
comet dust comae is expected to be dominated by dust grains in the
sub-micrometer to micrometer size range \citep{kolokolova04-comets2}, which
scatter optical light differently than centimeter-sized particles.

First we take the case in which the particles photometrically behave
like comet nuclei, i.e., they have a very low albedo, and have a phase
angle behavior like a macroscopic object and not like small dust
grains.  We refer to this case as the ``dusty case,'' but, just like
comet nuclei, these model particles may be internally icy.  We adopt
the geometric albedo ($A_p=0.049$) and phase function ($-2.5\log{\Phi}
= 0.046\theta$~mag for $\theta$ in units of degrees) of Hartley 2's
nucleus \citep{li12-thisissue}.  The geometric albedo is defined as the ratio of
the energy scattered from the object toward a phase angle of 0\degr{}
to that scattered from a white Lambertian disk with the same cross
section \citep[cf.][for this and other relevant albedo
  definitions]{hanner81-albedo}.

In Section~\ref{sec:flux-dist}, we found that individual particle
fluxes range from \fnorm$=1\times 10^{-13}$ to $4.5\times 10^{-11}$
W~m$^{-2}$~\micron$^{-1}${}, where \fnorm{} is the particle flux
normalized to a distance of 700~km.  To convert between flux and cross
section, we use the formula
\begin{equation}
  F_\lambda = A_p \Phi(\theta) \sigma
  \frac{S_{\lambda,\odot}}{r_h^2}\frac{1}{\Delta^2}
  \label{eq:flux}
\end{equation}
where $F_\lambda$ is the particle flux in units of
W~m$^{-2}$~\micron$^{-1}$, $\sigma$ is the cross-sectional area of the
particle (cm$^2$), $S_{\lambda,\odot}$ is the solar flux density at
1~AU (W~m$^{-2}$~\micron$^{-1}$), $r_h$ is the heliocentric distance
of the particle (1.064~AU), and $\Delta$ is the spacecraft-particle
distance (cm).  For the HRIVIS and MRI CLEAR1 filters, we use a solar
flux density of 1471 and 1435 W~m$^{-2}$~\micron$^{-1}${},
respectively \citep{klaasen12}.  To convert from cross section to
effective radius, we assume a spherical geometry: $\sigma=\pi a^2$.
Altogether, assuming a comet nucleus-like photometric behavior, the
effective radii of the particles range from 10 to 221~cm.  The largest
of these particles (4~m diameter) are just over the resolution of the
HRIVIS reconstructed frames (about 3~m at a distance of 700~m).  We
consider these sizes to be an upper limit to the true particle sizes.
For a lower-limit estimate, we consider the case where the particle
scattering function is similar to the icy satellite Europa: $A_p =
0.67$ and $\Phi = 1.0 - 0.01 \theta + 2.2\times 10^{-5} \theta^2$
\citep{buratti83, grundy07}.  At a phase angle of 80\degr, Europa's
phase function is 0.34.  For this parameter set the effective radii
are 0.8 to 17~cm.

Following Eq.~\ref{eq:flux}, where flux is proportional to
cross-sectional area, we can compute the total observed particle cross
section and add it to Table~\ref{tab:extrapolate}.  We have assumed
our icy particle case, using the photometric parameters of Europa.  To
instead use our dusty case, multiply the cross sections in the table
by 158.

Our flux distributions imply a very steep size distribution.  Assuming
a spherical geometry, the differential size distribution is
\begin{equation}
  \frac{dn}{da} = 2 N_0 F_1^{\alpha + 1}a^{2\alpha + 1},
\end{equation}
where $F_1$ is the flux from a 1~cm radius particle in units of
W~m$^{-2}$~\micron$^{-1}${}, and particle radius $a$ is in units of cm.  The constant $N_0$
is the solution to the equation:
\begin{equation}
  N = N_0 \int_{F_{\lambda,{\rm min}}}^{F_{\lambda,{\rm max}}}\frac{dn}{dF}dF,
\end{equation}
using the appropriate values from Table~\ref{tab:extrapolate}.  Our
low-flux power-law slope ($\alpha=-2.85$) corresponds to a size
distribution proportional to $a^{-4.7}$.  This slope is steeper than
the many size distribution estimates of small through large dust
grains ($a\sim1$~\micron{} to 1~mm with slopes near $-4$ to $-3$)
based on grain thermal emission \citep{lisse98, harker02-hb,
  harker11}, grain dynamics \citep{fulle04, reach07, kelley08,
  vaubaillon10}, and dust flux monitors on spacecraft
\citep{mcdonnell87, green04}.  Specifically for Hartley~2,
\citet{bauer11} estimate the comet's overall size distribution to
follow a power-law slope of $-4.0\pm0.3$, derived by comparing
$R$-band and WISE 12 and 22~\micron{} fluxes.  \citet{epifani01} fit
an ISOCAM 15~\micron{} image of the comet taken about 10~days after
perihelion with a dust dynamical model.  They report a time-averaged
power-law slope of $-3.2\pm0.1$, but inspection of their Fig.~10
suggests $-3.8$ is more appropriate (their best-fit slopes never fall
above $-3.5$).  These slopes are shallower than our value of $-4.7$,
but there is no requirement that they be the same.

To better understand the size, and thereby the composition, of the
large particles, we investigate the largest particle that may be
lifted from the nucleus, $a_{crit}$.  This parameter is estimated by
comparing the force of gravity to the gas drag force at the surface of
the comet.  \citet{meech04-activity} integrated the equation of motion
for spherical particles ejected from a spherical nucleus and found
\begin{equation}
  a_{crit} = \frac{9 \mu \mH Q v_{th}}{64 \pi^2 \rho_p \rho_N R_{N}^3 G},
\end{equation}
where $\mu$ is the atomic weight of the gas (amu), \mH{} is the mass
of hydrogen (g), $Q$ is the gas production rate (s$^{-1}$), $v_{th}$
is the mean thermal expansion speed of the gas (cm~s$^{-1}$), $\rho_p$
is the density of the particle (g~cm$^{-3}$), $\rho_N$ is the density
of the nucleus (g~cm$^{-3}$), $R_N$ is the radius of the nucleus (cm),
and $G$ is the gravitational constant (cm$^3$~g$^{-1}$~s$^{-2}$).
With the shape model of Comet Hartley~2, \citet{thomas12-hartley2}
estimate the surface gravity of the nucleus to be $a_N =
0.0019-0.0044$~cm~s$^{-2}$, which includes the rotation state of the
nucleus.  Therefore, we re-write the equation from
\citet{meech04-activity} to use the gravitational acceleration at the
nucleus
\begin{equation}
a_{crit} = \frac{3 \mu \mH Q v_{th}}{16 \pi \rho_p a_N R_{N}^2}.
\label{eq:acrit}
\end{equation}
The bulk material density for dust is $\sim3$~g~cm$^{-3}$ and for ice
is 1.0~g~cm$^{-3}$.  For the dusty case, we will consider porous
aggregates of dust with a total density of 0.3~g~cm$^{-3}$ (i.e., 90\%
vacuum).  For the icy case, we will at first assume 1.0~g~cm$^{-3}$,
but later consider porous aggregates with $\rho_p=0.1$~g~cm$^{-3}$.

The total water production rate of Comet Hartley 2 near closest
approach has been estimated via several methods to be
$Q(\water)\approx0.7-1.2\times 10^{28}$~s$^{-1}$ \citep{ahearn11,
  combi11-h2, dellorusso11, meech11-epoxi, mumma11-h2, knight12-thisissue}.
Yet, the coma contains a significant amount of water ice
\citep{ahearn11, protopapa11}, which could be supplying a large
fraction of the water vapor around the comet.  Moreover, the water
production rate is not uniformly distributed over the surface
\citep{ahearn11}.  We can account for these observations by
multiplying the water production rate by the ratio $f_{surface} /
f_{active}$, where $f_{surface}$ is the fraction of the water vapor
produced at the surface of the nucleus, and $f_{active}$ is the areal
fraction of surface that is active.  For illustrative purposes, we
will assume that 1\% of the water vapor is produced from 10\% of the
surface.  The remaining 99\% of the water vapor sublimates from the
icy grain halo.

For a nucleus water production rate of $Q(\water) f_{surface} /
f_{active} = 10^{27}$~s$^{-1}$, $v_{th}=0.5$~km~s$^{-1}$, Hartley~2's
mean radius (0.58~km), mean surface gravity, and assuming a particle
density of pure ice ($\rho_p=1$~g~cm$^{-3}$), we find $a_{crit}=8$~cm.
If we instead take \coo{} as the driving gas, with a production rate
of $10^{27}$~s$^{-1}$ \citep{ahearn11}, $f_{surface}=100\%$, and
$f_{active}=10\%$, then $a_{crit}$ becomes 200~cm for solid ice
spheres.  Based on this exercise, the icy model size estimates,
$a\leq17$~cm, are reasonable.

If instead of icy particles, we assume nucleus-like particles with a
density of 0.3~g~cm$^{-3}$, our $a_{crit}$ estimates increase to 28~cm
(\water) and 670~cm (\coo).  Compared to our size estimates of
$a\leq210$~cm, dark, dusty particles are plausible if \coo{} is the
driving gas; water can be made consistent if $f_{surface}/f_{active}$
is increased to 1.0.

Aggregate particles are more easily lifted by gas drag due to their
larger surface area per mass.  \citet{nakamura98} found that the drag
force for an aggregate is approximately the same as the drag force on
an area-equivalent sphere (with an error of less than 40\% in the
large aggregate limit).  Since our particle radii are based on the
observed flux, which is proportional to the cross-sectional area, our
radii are already defined by area-equivalent spheres.  Therefore, by
revising our particle density we can use Eq.~\ref{eq:acrit} to
estimate $a_{crit}$ for aggregates.

We assembled model particles using a ballistic particle-cluster
aggregate (BPCA) method \citep{meakin84}.  As the size of the
aggregate grows beyond a few thousand monomers the density
asymptotically approaches 10\% of the bulk material density.  Thus,
for a refractory material with a bulk density near 3~g~cm$^{-3}$, a
centimeter-sized BPCA particle would have a density of
0.3~g~cm$^{-3}${}.  The $a_{crit}$ estimates will be the same as in
our nucleus-like case above.

Treating the large particles as aggregates rather than solid spheres
better agrees with the HRIIR spectra of the coma.  \citet{ahearn11}
and \citet{protopapa11} studied the water ice absorption features and
found that they are most consistent with icy aggregates with monomer
radii $\approx1$~\micron.  An icy BPCA particle would have a density
of 0.1~g~cm$^{-3}$. The lowered density for icy aggregates increases
$a_{crit}$ by a factor of 10, giving us a healthy margin for launching
large icy particles off the surface of the nucleus, even where water
sublimation is driving the activity.

In summary, the dusty particle case produces very large particle
estimates (up to 2~m in radius) that are just at the resolution limit
of the HRIVIS instrument.  Gas expansion from \coo{} is sufficient to
lift these large particles from the surface of the comet if they have
a comet-like density of $\approx0.3$~g~cm$^{-3}$.  The water-ice case
produces particle estimates up to $\approx20$~cm in radius, which are
easily lifted from the nucleus by water or \coo{} expansion.

\section{Spatial distribution and origin}
\begin{figure}
  \ifincludegraphics
  \begin{center}
    \includegraphics[width=1.0\columnwidth]{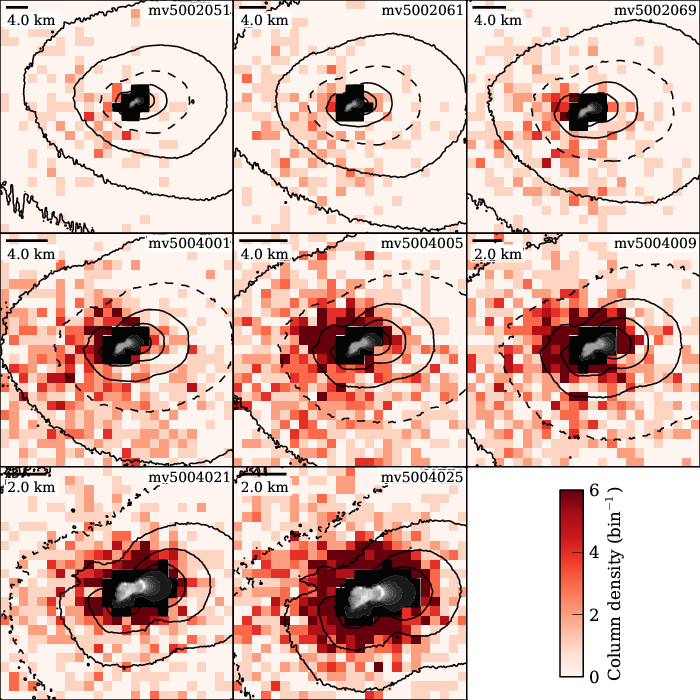}
  \end{center}
  \fi

  \caption[MRI number maps 1]{Contours of the total coma surface
    brightness superimposed over particle column density for 501 and
    121 ms MRI images mv5002051 through mv5004025.  The contours are
    spaced at factor of 2 intervals, and the dashed line is 0.008
    W~m$^{-2}$~\micron$^{-1}$~sr$^{-1}$.  The coma image was smoothed
    with a Gaussian kernel function (7~pixel FWHM) before the contours
    were created.  Only particles with $F_\lambda > F_{\lambda, {\rm
        min}}$ are considered, and the particle bins are $38\times38$
    pixels in size.  The region closest to the nucleus is masked from
    the analysis, and has been replaced with each epoch's image of the
    nucleus and inner-most coma.  In all images, the projected
    velocity of the spacecraft is toward the bottom, and the sunward
    direction is approximately to the
    right.  \label{fig:mri-numbermaps-1}}
\end{figure}

\begin{figure}
  \ifincludegraphics
  \begin{center}
    \includegraphics[width=1.0\columnwidth]{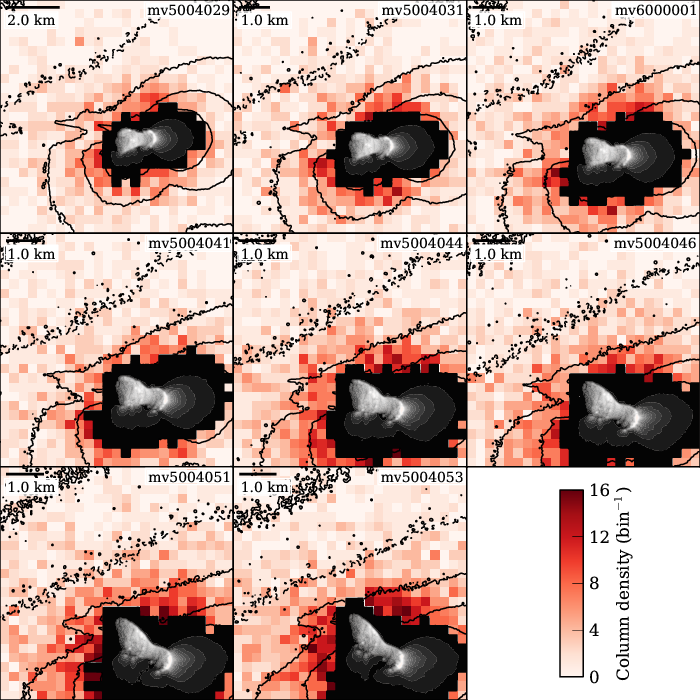}
  \end{center}
  \fi

  \caption[MRI number maps 2]{Same as Fig.~\ref{fig:mri-numbermaps-1},
    but for images mv5004029 through
    mv5004053.\label{fig:mri-numbermaps-2}}
\end{figure}

\begin{figure}
  \ifincludegraphics
  \begin{center}
    \includegraphics[width=1.0\columnwidth]{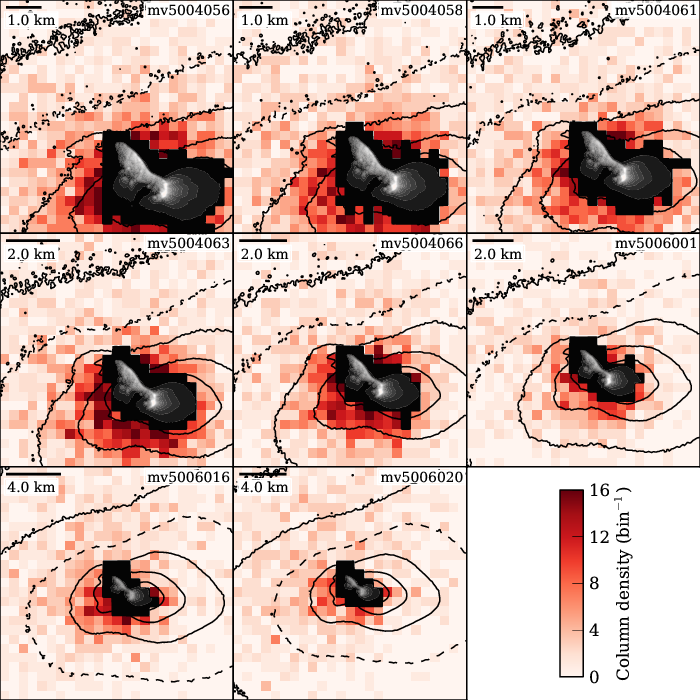}
  \end{center}
  \fi

  \caption[MRI number maps 3]{Same as Fig.~\ref{fig:mri-numbermaps-1},
    but for images mv5004056 through
    mv5006020.\label{fig:mri-numbermaps-3}}
\end{figure}

\begin{figure}
  \ifincludegraphics
  \begin{center}
    \includegraphics[width=1.0\columnwidth]{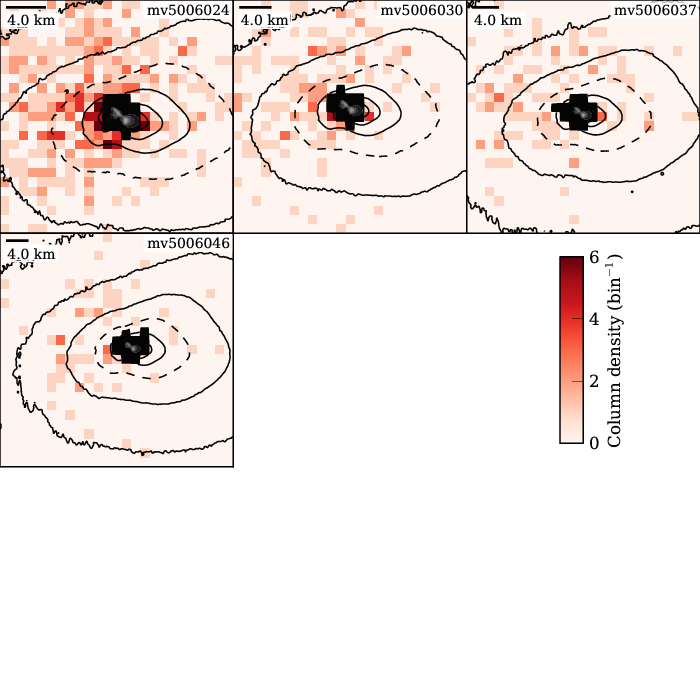}
  \end{center}
  \fi

  \caption[MRI number maps 4]{Same as Fig.~\ref{fig:mri-numbermaps-1},
    but for images mv5006024 through
    mv5006046.\label{fig:mri-numbermaps-4}}
\end{figure}

Mapping the spatial distribution of the particles gives clues to their
origins and dynamics.  In
Figs.~\ref{fig:mri-numbermaps-1}--\ref{fig:mri-numbermaps-4}, we plot
the column density of particles and total coma surface brightness
contours for all 501 and 121~ms MRI images listed in
Table~\ref{tab:fits}.  The column density images were derived from our
final photometry lists for $F_\lambda > F_{\lambda, {\rm min}}$,
binned onto a $38\times38$ pixel grid.  Inspection of the figures
reveals that the coma and the particles have different spatial
distributions.  The particles are biased to the anti-sunward direction
on scales $>2-4$~km, whereas the coma distribution is dominated by the
jets.  This asymmetry is especially apparent in the strong sunward
jets, where the particle density is lower than in other regions of
similar surface brightness.  It is not an observational bias; we have
masked those regions close to the nucleus where the column density and
jet morphology interfere with the PSF fitting process.  Instead, the
low particle density in the sunward jets can be accounted for by
particle dynamics.  We consider three dynamical processes that could
be affecting the particle distribution: (1) the rotation of the
nucleus; (2) solar radiation pressure; and (3) a rocket effect from
sublimating ice.  Hydrodynamic flows from the strong gas production
rates and asymmetry in outgassing may also play a role in the particle
dynamics, but this analysis is outside the scope of this paper.

\subsection{Radial expansion and nucleus rotation state}
Rotation of the nucleus has a direct impact on the spatial
distribution of particles.  To estimate this effect, we must first
recognize that the particle outflow speeds are low.  In order for the
particles to be seen as point sources in HRIVIS images at closest
approach, their speeds must be lower than $\approx
3~\mbox{m}/0.376~\mbox{s} = 8$~m~s$^{-1}$.  A lower constraint is
computed by \citet{hermalyn12} based on the 3D positions of the
particles.  They find 0.5--2~m~s$^{-1}$ to be more typical (note that
these speeds are not necessarily radial).  At such low speeds, the
large particles take $\gtrsim10^3$~s to reach 2~km from the nucleus.
In contrast, small dust grains move much more quickly with outflow
speeds expected to be of order 100~m~s$^{-1}$.  The fine dust reaches
2~km in as little as 20~s.

The positions of the major jets are governed by the rotation of the
long axis about the angular momentum vector, with a period of 18.4~h
near closest approach; the long axis is inclined to the angular
momentum vector by 81\degr{} \citep{belton12-hartley2}.  Taking
1~m~s$^{-1}$ as the particle outflow speed from the surface, the
nucleus will have rotated 10\degr{} by the time the particles have
traveled 2~km.  So, the rotation state has a minor consequence on the
distribution of particles at 2~km, and their spatial distribution
should be closely related to their source regions.  This observation
and the assumption of radial motion suggests that the strong sunward
jets are not the primary source of the large particles, but instead
they are ejected from along the long-axis of the nucleus.  The jets
pointed towards the bottom of the images in
Fig.~\ref{fig:mri-numbermaps-2} would be the next likely source region
for particles.  There is also a large population of particles towards
the top of Figs.~\ref{fig:mri-numbermaps-1} and
\ref{fig:mri-numbermaps-2}, but they do not have an apparent source
region (i.e., this side of the nucleus does not appear to be as active
as the other regions).  However, we note that not all potential
sources are apparent in the MRI images.  For example, the apparent
water jet seen in Fig.~5 of \citet{ahearn11}, which points to the top
right of Figs.~\ref{fig:mri-numbermaps-3} and
\ref{fig:mri-numbermaps-4}, has no clear optical counterpart but may
also contribute to the large particle production (n.b., this water jet
does not appear to contain ice, and therefore is unlikely to be a
source of large icy particles).

Taking a lower ejection speed of 1~cm~s$^{-1}$, the nucleus can rotate
three times before particles travel 2~km.  Thus on this length scale, and
in the absence of any other perturbing forces, the particles would
have a spatial distribution correlated with the activity profile of
the nucleus.  Because the CO$_2$ and H$_2$O gas production rates and
the optical light curve peak when the small end is pointed towards the
Sun \citep{ahearn11}, in the absence of other forces the large
particle density should peak in the solar direction, which is not
observed.

\subsection{Solar radiation pressure}\label{sec:qpr}
In principle, solar radiation pressure could redistribute large
particles into the anti-sunward direction.  The acceleration from
radiation, $a_{rad}$, in units of cm~s$^{-2}$ is \citep{burns79}
\begin{equation}
  a_{rad} = \frac{Q_{pr}S_\sun\sigma}{c m r_h^2},
  \label{eq:radiation}
\end{equation}
where $Q_{pr}$ is the radiation pressure efficiency factor, $S_\sun$
is the integrated flux density ($1.361\times 10^{6}$
erg~cm$^{-2}$~s$^{-1}$ at 1~AU), $\sigma$ is the geometrical
cross section of the particle (cm$^2$), $c$ is the speed of light
(cm~s$^{-1}$), $m$ is the mass of the particle in question, and $r_h$
is the heliocentric distance (AU).  We have already dropped all
velocity dependent terms from $a_{rad}$ \citep[cf.][]{burns79}.  It is
common to express radiation pressure with the parameter $\beta$
defined as the ratio of the force of solar radiation pressure to the
gravitational force from the Sun,
\begin{equation}
  \beta \equiv \frac{F_{rad}}{F_{grav}} = \frac{5.7\times10^{-5} Q_{pr}}{\rho_p a}.
\end{equation}

The radiation pressure efficiency is 1 for perfectly absorbing,
isotropically emitting spheres.  For our particles, we again take the
icy particle case and derive $Q_{pr}$ from the phase function and
geometric albedo \citep{vandehulst57}: $Q_{pr} \approx 1 - A_B
\overline{\cos{\alpha}}$, where $A_B$ is the Bond albedo, and
$\overline{\cos{\alpha}}$ describes the anisotropy of the scattered
light, where $\alpha=180 - \theta$ is the scattering angle.  The
relationship between the Bond albedo and geometric albedo is $A_B =
A_p \int_0^\pi \Phi(\alpha)\sin{\alpha}d\alpha$
\citep{hanner81-albedo}.  Altogether, we compute $Q_{pr}=1.58$.  For
the mass, we again assume two cases for particles with $a=10$~cm: (1)
solid spheres with the density of ice, 1~g~cm$^{-3}$; and (2) compact
aggregates of ice (1~\micron{} radius monomers) with an overall
particle density of 0.1~g~cm$^{-3}$.  For the latter case, we note
that detailed calculations will be required to understand how albedo
and the anisotropy of scattering are affected by the complex aggregate
shape, and we reserve this investigation for future work.  We compute
accelerations of $4.8\times 10^{-6}$ and $0.5\times 10^{-6}$
cm~s$^{-2}$ for 10~cm solid and BPCA aggregate ice particles,
equivalent to $\beta=9\times10^{-6}$, and $1\times 10^{-6}$.  For
comparison, comet dust trails are comprised of grains with
$\beta\lesssim10^{-3}$ \citep{sykes92, reach07}.

In Figs.~\ref{fig:mri-numbermaps-1}--\ref{fig:mri-numbermaps-4},
particles are found out to the image edges in the sunward direction.
However, the sunward/anti-sunward asymmetry is clear on 2--4~km length
scales.  In the rest frame of the comet, the relationship between
turnaround distance ($d$), ejection speed ($v_{ej}$), and acceleration
from radiation pressure is $d=v_{ej}^2 / 2 / a_{rad}$.  Solving for
ejection speed $v_{ej}$ yields
\begin{equation}
  v_{ej}^2 = \frac{3 d Q_{pr} S_\sun}{2 c r_h^2 \rho_p a} =
  6.0\times10^{-5} \frac{d Q_{pr}}{\rho_p a},
\end{equation}
for $v_{ej}$ in units of cm~s$^{-1}$, $d$ in cm, $\rho_p$ in
g~cm$^{-3}$, and $a$ in cm.  For our icy particle case $v_{ej}=4.3-6.1
(\rho a)^{-1/2}$~cm~s$^{-1}$ (4--6~cm~s$^{-1}$ for a 10~cm aggregate,
and 1--2~cm~s$^{-1}$ for solid ice).  If particles are truly ejected
at these speeds, solar radiation pressure would take $\gtrsim10^6 \rho
a$~s to accelerate the particles to the $\gtrsim50$~cm~s$^{-1}$ speeds
measured by \citet{hermalyn12}.  This result also implies that the
dominant velocity component for particles more than a few kilometers
from the nucleus would be distinctly in the anti-solar direction, yet
only a weak asymmetry in the velocity is observed \citep{hermalyn12}.
Therefore, radiation pressure does not govern the dynamics for our icy
particle cases.  The same conclusion is reached for the dusty particle
case with a nucleus-like density ($Q_{pr}=1.00$,
$\rho_p=0.3$~g~cm$^{-3}${}): $v_{ej}=6.3-8.9a^{-1/2}$~cm~s$^{-1}$ or
0.6--0.9~cm~s$^{-1}$ for a 100~cm particle.

If we instead assume an ejection speed of 1~m~s$^{-1}$ for a 10~cm
particle, their densities must be
$\rho_p=2-4\times10^{-4}$~g~cm$^{-3}${} in order to be turned around
by $d=2-4$~km.  Thus, if the particles are very fluffy aggregates,
they may be ejected at larger speeds, and radiation pressure will be
able to redistribute them into the anti-solar direction, but they
could still have instantaneous velocities distributed about the
anti-solar vector.

\subsection{Rocket effect}\label{sec:rocket}
The rocket effect is the acceleration of the particles due to the
sublimation of water ice.  For a spherical geometry with radial
outflow,
\begin{equation}
  a_{rocket} = \frac{3 \mu \mH Z v_{th} f_{ice}}{4 \rho_p a} =
    1.1\times10^{-18}\frac{Z}{\rho_p a},
\end{equation}
where $Z$ is the sublimation rate of the particle
(molec~cm$^{-2}$~s$^{-1}$), $\mu$ is the molecular weight of the
sublimating ice (18~u for water), \mH{} is the mass of hydrogen (g),
and $f_{ice}$ is the ice fraction of the particle (all remaining
parameters are in cgs units).  Following \citet{reach09}, we define
the dimensionless rocket effect parameter $\alpha$ as the ratio of the
force from the sublimation mass-loss, $F_{rocket}$, to the force of
gravity from the Sun, $F_{grav}$:
\begin{equation}
  \alpha \equiv \frac{F_{rocket}}{F_{grav}} = 2.238\times10^{26}
  \frac{3 \mu \mH Z v_{th} f_{ice}}{4 G M_\sun \rho_p a},
  \label{eq:rocket}
\end{equation}
where $G$ is the gravitational constant (cm$^3$~g$^{-1}$~s$^{-2}$),
and $M_\sun$ is the mass of the Sun (g).  The rocket effect parameter
should have a heliocentric distance dependency, since the product $Z
v_{th}$ does not necessarily vary as $r_h^{-2}$, but
Eq.~\ref{eq:rocket} will serve as a good approximation for small
$\Delta r_h$.  The $\alpha$ parameter is analogous to the $\beta$
parameter for dust; both parameters quantify a force directed away
from the Sun in fractions of the solar gravitational force.

Assuming no losses from scattering or thermal emission, the maximum
water sublimation rate from ice ($Z_{max}$) can be computed from the
solar flux density, particle absorption efficiency ($Q_{abs}$), and
the latent heat of sublimation for water ice ($L$),
\begin{equation}
  Z_{max} = \frac{S_\sun Q_{abs} N_A}{r_h^2 L}
\end{equation}
where $N_A$ is Avogadro's number, and $L$ ranges from
$5.0\times10^{11}$ to $5.1\times10^{11}$~erg~mol$^{-1}$ for
temperatures from 100 to 300~K \citep{murphy05}.  For large compact
aggregates (i.e., porous spheres) and solid particles
$Q_{abs}\approx1$, but it is larger for fluffy aggregates since they
have light scattering properties more like a collection of monomers,
rather than a single solid particle \citep{kolokolova07}.  This last
comment aside, $Z_{max}$ becomes $1.4\times10^{18}$
molec~cm$^{-2}$~s$^{-1}$ for $T=300$~K.  To account for scattering,
$Z_{max}$ will scale with $1 - A_B$, where $A_B$ is the Bond albedo.
In \S\ref{sec:qpr} we computed $A_B=0.84$ for our icy particle model.
Scattering reduces the $Z_{max}$ of the icy model to
$Z_{max,icy}=2.2\times10^{17}$ molec~cm$^{-2}$~s$^{-1}$.  For our
dusty model, $A_B=0.013$ and $Z_{max,dusty}\approx Z_{max}$.

For comparison, consider the model of \citet{beer06} for the
sublimation rate of icy particles.  They employ spherical grains,
heated by absorption of sunlight and cooled by thermal emission and
sublimation, and computed the absorption and emission efficiencies
with Mie scattering and effective medium theory considering both pure
ice grains, and ``dirty-ice'' grains, i.e., ice mixed with a generic
absorber (dust).  Mixing dust with the ice has a significant effect on
the ice equilibrium temperature, but the effect is not a strong
function of the ice-to-dust mass ratio (they tested $m_{ice} /
m_{dust} =0.9$ and 0.5).  Dust should also affect the overall albedo
of the particles, but they do not report this parameter.  In their
Figs.~8 and 9 they present computed grain lifetimes, defined as the
time a grain takes to completely sublimate.  Their lifetimes for
$r_h=1.09$~AU are the best examples for our scenario.  For $a >
0.01$~cm, pure ice grains have lifetimes of $\tau_{pure}=1\times
10^{10}a$~s for particle radius $a$ measured in cm.  Dirty-ice grains
have a much shorter lifetime: $\tau_{dirty}=1\times 10^{5}a$~s.  Since
the large particles spend most of their lifetime larger than 0.01~cm,
we can transform their lifetimes into sublimation rates:
\begin{equation}
  Z = \frac{4 \rho_p a}{3 \mu \mH \tau f_{ice}}.
\end{equation}
The \citet{beer06} model sublimation rates for large particles are
$Z_{dirty}=4\times10^{17}$ molec~cm$^{-2}$~s$^{-1}$, and
$Z_{pure}=4\times10^{12}$ molec~cm$^{-2}$~s$^{-1}$.  These values are
comparable to or less than our maximum sublimation rates.

Taking $Z_{max,icy}$, a 10~cm radius particle is accelerated at a rate
$a_{rocket} = 0.024$~cm~s$^{-2}$ away from the Sun, yielding a
rocket parameter $\alpha=0.042$, i.e., the rocket effect is 4\% the
force of solar gravity.  For $Z=Z_{max}$, we compute $\alpha=0.27$.
Unlike radiation pressure, the rocket effect is potentially very
strong.

Acceleration from sublimation will distribute the particles in the
anti-solar direction.  Following our method for radiation pressure, we
can constrain the particle ejection speed with the implied sublimation
rate
\begin{equation}
  v_{ej}^2 = 2.2\times10^{-18}\frac{Z d}{\rho_p a}.
\end{equation}
Again, adopting 2--4~km as our typical turnaround distance, and taking
$Z=Z_{max,icy}$, we find ejection speeds of $310-440
a^{-1/2}$~cm~s$^{-1}$ for our solid ice case ($990-1400
a^{-1/2}$~cm~s$^{-1}${} for icy aggregates).  These ejection speeds
are higher than the instantaneous speeds measured by
\citet{hermalyn12}, but the two speeds do not need to agree since one
is at/near the surface and the other is out in the coma.  For
$Z=Z_{max}$, the resulting ejection speeds are increased:
$v_{ej}=780-1100 a^{-1/2}$~cm~s$^{-1}${} (solid ice) and $2500-3500
a^{-1/2}$~cm~s$^{-1}${} (icy aggregates).  However, in the above
analysis we have assumed that only the sunlit hemisphere is
sublimating.  By distributing the sublimation across more of the
surface, we can decrease the implied ejection speeds to
10--100~cm~s$^{-1}$, similar to the speed measured in the coma by
\citet{hermalyn12}.  Therefore, we conclude that a sublimation rate
\textit{excess} on the sunlit hemisphere of order $10^{17}$
molec~cm$^{-2}$~s$^{-1}${} readily describes the sunward/anti-sunward
particle asymmetry.  Detailed simulations will be needed to fully
account for the observed distribution of particle velocities.

\section{Mass and water production rate}\label{sec:mass-water}
Table \ref{tab:extrapolate} includes the total particle mass, based on
our icy photometric model and a particle density of 1~g~cm$^{-3}$.
For the dusty model and $\rho_p=0.3$~g~cm$^{-3}$, the masses are
increased by a factor of 672.  With our preferred flux lower limit of
0.1--$1.0\times10^{-12}$, the particle masses correspond to $0.03-0.10
M_N$ (icy), $23-70 M_N$ (dusty), where $M_N=2.4\times10^{11}$~g is the
mass of the nucleus, assuming a 0.3~g~cm$^{-3}$ density
\citep{thomas12-hartley2}.  It is clear that the dusty case is
impossible.  The only way we can reduce the estimated total mass for
these dark particles is by decreasing their densities to well below
$10^{-3}$~g~cm$^{-3}$.  Therefore, we do not favor the dusty case, but
it does remain as a possible interpretation.  Given that the total
mass lost from the comet per orbit is of order 1\% of the nucleus
\citep{thomas12-hartley2}, even the solid ice particles may be too
massive.  Porous ice particles (e.g., $\rho_p=0.1$~g~cm$^{-3}$) should
be considered the most likely case.

Icy particles will begin sublimating as soon as they are released from
the nucleus and warmed by insolation.  In \S\ref{sec:rocket}, we
computed the rocket effect on the particles due to water ice
sublimation and concluded that a sublimation rate excess of $10^{17}$
molec~cm$^{-2}$~s$^{-1}${} on the particle's sunlit hemisphere can
describe the observed sunward/anti-sunward asymmetry in particle
column density.  We also computed the maximum sublimation rate, based
on energy balance between absorbed solar radiation and sublimation.
We apply this latter value, $Z_{max,icy}$ to all of the large
particles and compute water production rates.  The total large
particle water production rate within a 20.6~km radius aperture is
limited to $<0.6-5\times10^{25}$~molec~s$^{-1}$, which is $<0.1-0.5$\%
of the total water production rate of the comet
($\approx1\times10^{28}$ molec~s$^{-1}$).  We can change the water
production rate by assuming a different photometric model as
$Q\propto(1 - A_B)(A_p\Phi)^{-1}$.  For our nucleus-like case, we find
$Q({\rm H_{2}O})<16-80\%$ of the total water production rate, but,
unless the particles are fluffy aggregates with
$\rho\lesssim10^{-3}$~g~cm$^{-3}$, we rule out this case based on
their mass.  The icy particle case yields our best estimate of the
water production rate, $Q({\rm H_{2}O})<(1-5\times10^{-3})Q_{total}$.

\section{Comparison to other observations}\label{sec:radar}
\citet{harmon11} observed comet Hartley~2 with Arecibo S-band
($\lambda=12.6$~cm) radar at the end of October 2010, about a week
before \di's closest-approach.  In their average Doppler spectrum,
they observe a strong grain-coma echo, with a characteristic radial
velocity dispersion of 4~m~s$^{-1}$.  The velocity distribution is
asymmetric, with a range of $\approx -50$ to $+13$~m~s$^{-1}$ with
respect to the nucleus (negative velocities are away from the Earth).
The coma has a strongly depolarized echo, which indicates the radii of
the largest particles are well above the Rayleigh limit of
$\lambda/2\pi=2$~cm.  \citet{harmon11} suggest that there exists a
significant population of particles with decimeter sizes or larger.
We explore the possibility that the radar observations may be the
large particles imaged by \di.

The average radar cross section of the coma was 0.89~km$^2$.  From our
MRI observations, we derived a total icy particle cross section of
$\sigma=4\times10^{-4}$ to $3\times10^{-3}$~km$^2$ within 20.6~km from
the nucleus.  Our cross section is more than two orders of magnitude
smaller than the radar observed cross section.  However, their beam is
much larger than what the MRI can image when individual particles are
detectable.  We have found particles out to 40~km in
Fig.~\ref{fig:mri-extent}, but the Arecibo beam size is 32,000~km at
the distance of the comet.  With speeds of order 4~m~s$^{-1}$, and
lifetimes of order $10^5$~s (dirty ice), the large icy particle coma
would extend out to $\sim400$~km.  This estimate implies our census of
the large particles is incomplete by about a factor of $<400 / 20.9 =
20$, resulting in a total icy particle cross section of
$\lesssim0.01$~km$^2$, which still remains inconsistent with the radar
results.

Assuming for the moment that the radar cross section properly reflects
the total icy particle population we compute an upper limit to the
total water production rate of $<2\times 10^{27}$~molec~s$^{-1}${}.
The SWAN instrument on the \textit{SOHO} satellite observes Ly$\alpha$
emission over fields of view much larger than the radar beam size
($10^5$ km~pixel$^{-1}$).  These observations are a good point of
reference for a total water production rate that is sure to include
water produced by any large particles observed with Arecibo.
\citet{combi11-h2} measured a water production rate of 6--$9\times
10^{27}$~molec~s$^{-1}${} near perihelion \citep{combi11-h2}.  By this
estimate, it seems that the large particles could account for a
substantial fraction of the total water production rate of the comet.
Note, however, that the SWAN-based water production rates are on par
with those observed in 4000~km apertures and smaller.  Based on the
aperture sizes and water production rates listed in
Table~\ref{tab:qwater}, most of the water is produced close to the
nucleus, perhaps within a few tens or hundreds of kilometers,
suggesting that the Arecibo observed particles have a low water
sublimation rate, if any.

A dusty large particle population yields a cross section of
$\sigma=0.07-0.5$~km$^2$ (\S\ref{sec:sizes}), which is in better
agreement with the radar results, but suggests that the entire large
particle coma is within $\sim40--250$~km from the comet.  There is no
indication in Figs.~\ref{fig:mri-numbermaps-1} and
\ref{fig:mri-numbermaps-4} that the large particle coma is truncated
on this length scale.  Moreover, dusty particles may not have
sublimating ice.  Instead, they could fragment into finer particles.
If the large particles are dusty, they must fragment on
$\lesssim40--250$~km length scales in order to keep their total cross
section less than or equal to the observed radar cross section.  Just
based on the observed cross sections, we consider the dusty case to be
less likely than the icy case, but still find the icy case to be
lacking.  Of course, a coma of both icy and dusty particles is
possible.  More information on the composition and light scattering
properties, including radar wavelengths, will be needed to reconcile
the radar and MRI observations.

\section{Summary}
Comet Hartley 2 is surrounded by a coma of large particles with radii
$\gtrsim1$~cm.  With observations from \di{}, we measured their total
flux and flux distribution, based on photometry of individual
particles.  The flux distribution of these particles implies a very
steep size distribution with power-law slopes ranging from $-6.6$ to
$-4.7$.  We estimate that the particles account for 2--14\% of the
total flux from the near-nucleus coma.  The spatial distribution of
the particles is biased to the anti-sunward direction, as observed by
the spacecraft both pre- and post-closest approach.  Radial expansion
from the active areas of the rotating nucleus does not explain the
observed spatial distribution, even if the ejection speeds are very
low ($\sim1$~cm~s$^{-1}$).  Radiation pressure from sunlight cannot
redistribute them into the anti-sunward direction on small enough
length scales unless the particles have extremely low densities
($\sim10^{-4}$~g~cm$^{-3}$) or low radial ejection velocities
($\lesssim10$~cm~s$^{-1}$).  Low ejection velocities suggest there
should be a strong anti-sunward velocity component in the coma, but
this does not agree with the velocity distribution observed by
\citet{hermalyn12}.

We examined two possible particle compositions.  Our models were based
on the photometric properties of the nucleus of Hartley 2 (dusty case:
low albedo, 0.3~g~cm$^{-3}$) and the Jovian satellite Europa (icy
case: high albedo, 0.1--1.0~g~cm$^{-3}$), and serve as approximate
limiting cases.

The dusty case produces particle size estimates ranging from 10~cm to
2~m in radius, the largest of which is just at the limiting resolution
of \di{}'s HRIVIS camera at closest approach.  Such large particles
may be lifted off the nucleus by gas drag if \coo{} is the driving
gas.  Water is a plausible alternative if the water production rate
from the nucleus is at least $10^{27}$~molec~s$^{-1}$.  Based on the
dusty model, the total large particle cross section within 20.6~km
from the nucleus is 0.07-0.5~km$^{2}$, similar to the 0.89~km$^2$
radar cross section observed by \citet{harmon11}.  If these particles
are mini-nuclei, we estimate they account for 16--80\% of the comet's
total water production rate (within 20.6~km).  However, we can all but
rule out the dusty case based on total mass estimates of the large
particles, which are in excess of 10 nucleus masses for densities of
0.3~g~cm$^{-3}$.  Densities $\lesssim10^{-3}$~g~cm$^{-3}$ are required
to reduce the total mass of the particles to a few percent of the
nucleus, which is needed to keep the particle mass less than the total
mass lost from the comet per orbit \citep[2\% during the 2010
  apparition;][]{thomas12-hartley2}.

The icy case produces particle size estimates ranging from 1 to 20~cm
in radius.  These particles are easily lifted by water and \coo{} gas
drag.  Icy particles would sublimate as soon as they are heated by
sunlight.  If the particles have a net sublimation on their sunlit
sides, they would feel a rocket force that could easily distribute the
particles into the anti-sunward direction.  The sublimation rate
excess required for this redistribution is
$\lesssim10^{17}$~molec~cm$^{-2}$~s$^{-1}$, where the exact value
depends on the particle ejection velocities.  The cross section of icy
particles within 20.6~km is much smaller than the observed radar cross
section by two to three orders of magnitude.  The water production
rate of the \di{} observed particles is limited to $<0.1-0.5$\% of the
comet's total water production rate.  The total mass of the particles
for a density of 1~g~cm$^{-3}$ is 3--10\% the total mass of the
nucleus.  Thus, porous aggregates with $\rho_p\lesssim0.1$~g~cm$^{-3}$
should be considered more likely than solid ice particles.

We consider the icy case to be more likely than the dusty case for
three reasons: (1) the icy particles are more easily lifted by gas
drag; (2) we can account for the sunward/anti-sunward asymmetry in the
particle distribution if ice is sublimating on their sunlit sides; and
(3) the total large particle mass for the dusty case is much greater
than the total mass of the nucleus.  However, several details are
needed in order to test our hypothesis.  We need an improved icy
particle model that treats the large particles more like macroscopic
objects to better understand their water production rates (if icy) and
light scattering properties (dusty or icy).  We need to better
constrain the particle densities, which may rule out the dusty case.
A hydrodynamic analysis of the near-nucleus coma, especially around
the highly active small end, would improve our knowledge of the
dynamics of large particles.  A better understanding of the radar coma
that includes grains smaller than $\lambda/2\pi$ could help resolve
the discrepancy between our icy particle cross section and the
observed radar coma cross section.

If indeed icy with a high albedo, the large particles do not appear to
be the source of the comet's enhanced water production rate; although,
as discussed above, there is much work that can be done to refine this
conclusion.  We suspect that the small icy grains in the jets, as
observed in \textit{Deep Impact} IR spectra \citep{ahearn11,
  protopapa11}, are a significant source of water, and the primary
cause of the hyperactivity of Comet Hartley 2.

\section*{Acknowledgments}
The authors thank Lev Nagdimunov (UMD) for assistance in computing
cluster aggregate porosities, Adam Ginsberg for providing the
power-law fitting code, and Bj\"orn Davidsson and an anonymous referee
for helpful comments that improved this manuscript.

This work was supported by NASA's Discovery Program contract
NNM07AA99C to the University of Maryland and task order NMO711002 to
the Jet Propulsion Laboratory.          

This research made use of the PyRAF and PyFITS software packages
available at \url{http://www.stsci.edu/resources/software_hardware}.
PyRAF and PyFITS are products of the Space Telescope Science
Institute, which is operated by AURA for NASA.


\begin{table}
  \renewcommand\baselinestretch{1}
  \centering
  \caption[Data set]{HRIVIS and MRI images considered in this paper.
    Table columns are: $t-t_{enc}$, time relative to encounter; Exp.,
    exposure time; $\Delta$, spacecraft-comet distance; Scale, pixel
    scale at the distance of the comet; $\phi$, phase
    (Sun-comet-spacecraft) angle.}
  \label{tab:data}
  \scriptsize
  \begin{tabular}{lrrrrr}
    \hline\noalign{\smallskip}
    Image &
    $t - t_{enc}$ &
    Exp. &
    $\Delta$ &
    Scale &
    $\phi$
    \\
    &
    (s) &
    (ms) &
    (km) &
    (m) &
    (\degr)
    \\
    \hline\noalign{\smallskip}
    \multicolumn{6}{c}{HRIVIS} \\
    \hline\noalign{\smallskip}
    hv5004024 &  $-$48.4 &   375.5 &  915 &   1.8 & 79.6 \\
    hv5004025 &  $-$39.4 &   125.5 &  847 &   1.7 & 79.3 \\
    hv5004027 &  $-$29.1 &   375.5 &  781 &   1.6 & 79.0 \\
    hv5004028 &  $-$18.8 &   125.5 &  732 &   1.5 & 79.0 \\
    hv5004030 &   $-$9.8 &   375.5 &  704 &   1.4 & 79.1 \\
    hv5004031 &   $-$0.7 &   125.5 &  694 &   1.4 & 79.6 \\
    hv5004033 &      9.6 &   375.5 &  704 &   1.4 & 80.5 \\
    hv5004034 &     19.8 &   125.5 &  735 &   1.5 & 81.5 \\
    hv5004036 &     30.2 &   375.5 &  787 &   1.6 & 82.7 \\
    hv5004037 &     40.1 &   125.5 &  851 &   1.7 & 83.8 \\

    \hline\noalign{\smallskip}
    \multicolumn{6}{c}{MRI} \\
    \hline\noalign{\smallskip}
    mv5002051 & $-$414.0 &   500.5 & 5148 &  51.5 & 84.8 \\
    mv5002061 & $-$326.5 &   500.5 & 4082 &  40.8 & 84.4 \\
    mv5002069 & $-$254.4 &   500.5 & 3210 &  32.1 & 84.0 \\
    mv5004001 & $-$196.8 &   500.5 & 2523 &  25.2 & 83.4 \\
    mv5004005 & $-$167.0 &   500.5 & 2172 &  21.7 & 83.0 \\
    mv5004009 & $-$139.8 &   500.5 & 1858 &  18.6 & 82.6 \\
    mv5004012 & $-$119.9 &    40.5 & 1632 &  16.3 & 82.1 \\
    mv5004014 & $-$109.1 &   120.5 & 1513 &  15.1 & 81.8 \\
    mv6000000 & $-$100.3 &    40.5 & 1418 &  14.2 & 81.6 \\
    mv5004021 & $-$89.5 &   120.5 & 1303 &  13.0 & 81.2 \\
    mv5004023 & $-$79.6 &    40.5 & 1202 &  12.0 & 80.9 \\
    mv5004025 & $-$69.9 &   120.5 & 1107 &  11.1 & 80.5 \\
    mv5004027 & $-$59.0 &    40.5 & 1006 &  10.1 & 80.1 \\
    mv5004029 & $-$54.2 &   120.5 &  964 &   9.6 & 79.9 \\
    mv5004030 & $-$49.2 &    40.5 &  922 &   9.2 & 79.7 \\
    mv5004031 & $-$44.5 &   120.5 &  885 &   8.8 & 79.5 \\
    mv5004032 & $-$39.5 &    40.5 &  848 &   8.5 & 79.3 \\
    mv6000001 & $-$34.7 &   120.5 &  815 &   8.2 & 79.2 \\
    mv5004040 & $-$29.9 &    40.5 &  786 &   7.9 & 79.0 \\
    mv5004041 & $-$25.0 &   120.5 &  760 &   7.6 & 79.0 \\
    mv5004042 & $-$20.0 &    40.5 &  737 &   7.4 & 79.0 \\
    mv5004044 & $-$15.3 &   120.5 &  719 &   7.2 & 79.0 \\
    mv5004045 & $-$10.3 &    40.5 &  706 &   7.1 & 79.1 \\
    mv5004046 &  $-$5.5 &   120.5 &  697 &   7.0 & 79.3 \\
    mv6000002 &  $-$0.6 &    40.5 &  694 &   6.9 & 79.6 \\
    mv5004051 &   4.2 &   120.5 &  696 &   7.0 & 80.0 \\
    mv5004052 &   9.1 &    40.5 &  703 &   7.0 & 80.4 \\
    mv5004053 &  14.0 &   120.5 &  715 &   7.2 & 80.9 \\
    mv5004054 &  18.9 &    40.5 &  732 &   7.3 & 81.4 \\
    mv5004056 &  23.7 &   120.5 &  753 &   7.5 & 82.0 \\
    mv5004057 &  28.6 &    40.5 &  778 &   7.8 & 82.6 \\
    mv5004058 &  33.5 &   120.5 &  807 &   8.1 & 83.1 \\
    mv6000003 &  38.5 &    40.5 &  840 &   8.4 & 83.7 \\
    mv5004061 &  44.2 &   120.5 &  882 &   8.8 & 84.3 \\
    mv5004062 &  49.1 &    40.5 &  921 &   9.2 & 84.8 \\
    mv5004063 &  54.1 &   120.5 &  962 &   9.6 & 85.2 \\
    mv5004064 &  59.0 &    40.5 & 1004 &  10.0 & 85.7 \\
    mv5004066 &  68.9 &   120.5 & 1096 &  11.0 & 86.5 \\
    mv5006000 &  78.8 &    40.5 & 1192 &  11.9 & 87.2 \\
    mv5006001 &  88.7 &   120.5 & 1294 &  12.9 & 87.7 \\
    mv6000004 &  98.7 &    40.5 & 1399 &  14.0 & 88.2 \\
    mv5006011 & 109.5 &   120.5 & 1517 &  15.2 & 88.7 \\
    mv5006012 & 119.5 &    40.5 & 1627 &  16.3 & 89.1 \\
    mv5006016 & 142.4 &   500.5 & 1886 &  18.9 & 89.8 \\
    mv5006020 & 171.1 &   500.5 & 2219 &  22.2 & 90.4 \\
    mv5006024 & 207.1 &   500.5 & 2643 &  26.4 & 91.0 \\
    mv5006030 & 265.7 &   500.5 & 3345 &  33.5 & 91.6 \\
    mv5006037 & 335.2 &   500.5 & 4187 &  41.9 & 92.1 \\
    mv5006046 & 403.9 &   500.5 & 5023 &  50.2 & 92.4 \\
    \hline\noalign{\smallskip}

  \end{tabular}
\end{table}
\clearpage

\begin{SmallLongTable}
\begin{longtable}[c]{l rrr r@{$\,\pm\,$}l r r@{$\,\pm\,$}l r r}
  \caption[Power-law fits]{\textit{Online only table.}  All single
    power-law flux distributions derived from HRIVIS and MRI images.
    Table columns: $F_{\lambda}$, the minimum, maximum, and total
    particle fluxes in the fit; $N$, number of particles in the fit;
    $\alpha$, best-fit power-law slope and uncertainty; $P_{KS}$, the
    Kolmogorov-Smirnov probability that the distribution is a
    power-law; $F_T / F_{coma}$, the fraction of the total coma flux
    attributable to point sources with $F_{\lambda} > F_{\lambda, {\rm
        min}}$ in MRI full-frame data, excluding regions near the
    nucleus.  \label{tab:fits}}\\

  \hline\noalign{\smallskip}
  Image &
  $\Delta$ &
  \multicolumn{4}{c}{$F_{\lambda}$ $^{\mbox{a}}$} &
  $N$ &
  \multicolumn{2}{c}{$\alpha$} &
  $P_{KS}$ &
  $F_T / F_{coma}$
  \\\cline{3-6}
  &
  &
  min &
  max &
  \multicolumn{2}{c}{total} &
  \\
  &
  (km) &
  \multicolumn{4}{c}{($10^{-12}$ W~m$^{-2}$~\micron$^{-1}$)} &
  &
  \multicolumn{2}{c}{} &
  (\%) &
  
  \\
  \hline\noalign{\smallskip}
  \endfirsthead

  \caption*{\emph{Continued}}\\
  \hline\noalign{\smallskip}
  Image &
  $\Delta$ &
  \multicolumn{4}{c}{$F_{\lambda}$ $^{\mbox{a}}$} &
  $N$ &
  \multicolumn{2}{c}{$\alpha$} &
  $P_{KS}$ &
  $F_T / F_{coma}$
  \\\cline{3-6}
  &
  &
  min &
  max &
  \multicolumn{2}{c}{total} &
  \\
  &
  (km) &
  \multicolumn{4}{c}{($10^{-12}$ W~m$^{-2}$~\micron$^{-1}$)} &
  &
  \multicolumn{2}{c}{} &
  (\%) &
  
  \\
  \hline\noalign{\smallskip}
  
  \endhead
  
  \hline\noalign{\smallskip}
  \multicolumn{11}{l}{\emph{Continued on next page}} \\
  \endfoot
  
  \hline\noalign{\smallskip}

  \multicolumn{11}{p{5in}}{$^{\mbox{a}}$ The absolute calibration
    uncertainties are 5\% for HRIVIS, and 10\% for MRI.}
  
  \endlastfoot

 
\multicolumn{11}{c}{HRIVIS} \\*
\hline\noalign{\smallskip}
hv5004024 &  915 &  1.8 & 20.5 & \multicolumn{1}{r}{ 1853} & &  641 &  -3.06 &   0.09 & 0.3      & \nodata \\
hv5004025 &  847 &  1.5 & 31.6 & \multicolumn{1}{r}{ 3128} & &  986 &  -2.88 &   0.06 & $<$0.1   & \nodata \\
hv5004027 &  781 &  1.4 & 29.0 & \multicolumn{1}{r}{ 1388} & &  430 &  -2.86 &   0.08 & 0.6      & \nodata \\
hv5004028 &  732 &  1.5 & 33.1 & \multicolumn{1}{r}{ 2058} & &  685 &  -2.91 &   0.07 & 12.9     & \nodata \\
hv5004030 &  704 &  0.9 & 35.5 & \multicolumn{1}{r}{ 1222} & &  613 &  -2.67 &   0.07 & 16.8     & \nodata \\
hv5004031 &  694 &  1.5 & 40.6 & \multicolumn{1}{r}{ 1252} & &  419 &  -2.70 &   0.09 & 13.4     & \nodata \\
hv5004033 &  704 &  0.9 & 34.3 & \multicolumn{1}{r}{  918} & &  379 &  -2.67 &   0.08 & 15.2     & \nodata \\
hv5004034 &  735 &  1.5 & 31.2 & \multicolumn{1}{r}{ 1017} & &  374 &  -2.93 &   0.10 & 15.3     & \nodata \\
hv5004036 &  787 &  0.9 & 23.1 & \multicolumn{1}{r}{  902} & &  496 &  -2.94 &   0.09 & 8.6      & \nodata \\
hv5004037 &  851 &  1.5 & 21.2 & \multicolumn{1}{r}{ 1005} & &  425 &  -3.08 &   0.11 & 41.7     & \nodata \\

\hline\noalign{\smallskip}
\multicolumn{11}{c}{MRI sub-frame} \\*
\hline\noalign{\smallskip}
mv5004029 &  964 &  4.1 & 14.1 &   692 &     7 &  112 &  -4.08 &   0.29 & 38.5     & \nodata \\
mv5004031 &  885 &  4.5 & 13.4 &   660 &     7 &  114 &  -4.31 &   0.31 & 10.1     & \nodata \\
mv6000001 &  815 &  4.2 & 11.4 &   770 &     7 &  124 &  -4.03 &   0.27 & 8.2      & \nodata \\
mv5004041 &  760 &  3.9 & 16.6 &   836 &     7 &  135 &  -3.87 &   0.25 & 54.5     & \nodata \\
mv5004044 &  719 &  3.8 & 18.8 &   933 &     7 &  141 &  -3.44 &   0.21 & 3.0      & \nodata \\
mv5004046 &  697 &  3.6 & 12.9 &   631 &     5 &  124 &  -3.50 &   0.22 & 36.4     & \nodata \\
mv5004051 &  696 &  3.0 & 13.8 &   729 &     6 &  155 &  -3.22 &   0.18 & 13.5     & \nodata \\
mv5004053 &  715 &  3.1 & 21.3 &   553 &     5 &  110 &  -3.36 &   0.22 & 99.1     & \nodata \\
mv5004056 &  753 &  2.8 & 11.1 &   637 &     5 &  135 &  -3.34 &   0.20 & 35.7     & \nodata \\
mv5004058 &  807 &  3.0 &  9.0 &   437 &     4 &  116 &  -3.83 &   0.26 & 41.9     & \nodata \\

\hline\noalign{\smallskip}
\multicolumn{11}{c}{MRI 41 ms} \\*
\hline\noalign{\smallskip}
mv5004027 & 1006 &  5.5 & 16.4 &  1066 &     9 &  133 &  -4.05 &   0.26 & 67.3     &  0.0007 \\
mv5004030 &  922 &  6.7 & 17.8 &   884 &     9 &  108 &  -4.39 &   0.33 & 74.5     &  0.0006 \\
mv5004032 &  848 &  6.0 & 22.6 &  1773 &    12 &  189 &  -3.76 &   0.20 & 29.8     &  0.0012 \\
mv5004040 &  786 &  6.1 & 24.3 &  2053 &    13 &  252 &  -4.04 &   0.19 & 97.8     &  0.0019 \\
mv5004042 &  737 &  6.0 & 24.1 &  2483 &    15 &  321 &  -4.02 &   0.17 & 43.0     &  0.0032 \\
mv5004045 &  706 &  6.2 & 22.2 &  2747 &    15 &  331 &  -3.96 &   0.16 & 29.9     &  0.0040 \\
mv6000002 &  694 &  5.9 & 21.1 &  3406 &    17 &  390 &  -3.68 &   0.14 & 0.3      &  0.0046 \\
mv5004052 &  703 &  5.8 & 26.2 &  3100 &    16 &  374 &  -3.74 &   0.14 & 8.7      &  0.0044 \\
mv5004054 &  732 &  5.7 & 27.1 &  2822 &    15 &  309 &  -3.59 &   0.15 & 30.0     &  0.0035 \\
mv5004057 &  778 &  5.9 & 21.4 &  1977 &    13 &  208 &  -3.60 &   0.18 & 7.3      &  0.0023 \\
mv6000003 &  840 &  5.6 & 19.8 &  2029 &    13 &  235 &  -3.78 &   0.18 & 29.4     &  0.0018 \\
mv5004062 &  921 &  5.3 & 18.3 &  1359 &    10 &  188 &  -4.08 &   0.22 & 47.0     &  0.0011 \\
mv5004064 & 1004 &  4.9 & 15.7 &   922 &     9 &  149 &  -3.96 &   0.24 & 33.8     &  0.0008 \\

\hline\noalign{\smallskip}
\multicolumn{11}{c}{MRI 121 ms} \\*
\hline\noalign{\smallskip}
mv5004021 & 1303 &  2.4 & 10.6 &  2993 &    13 &  838 &  -3.95 &   0.10 & 8.4      &  0.0021 \\
mv5004025 & 1107 &  2.8 & 14.2 &  4374 &    16 & 1191 &  -3.87 &   0.08 & 3.2      &  0.0031 \\
mv5004029 &  964 &  3.4 & 13.9 &  6086 &    20 & 1291 &  -3.88 &   0.08 & 0.6      &  0.0039 \\
mv5004031 &  885 &  3.5 & 16.6 &  7945 &    23 & 1501 &  -3.87 &   0.07 & $<$0.1   &  0.0048 \\
mv6000001 &  815 &  4.1 & 19.7 &  7337 &    22 & 1262 &  -3.95 &   0.08 & 0.5      &  0.0050 \\
mv5004041 &  760 &  4.1 & 19.8 &  9078 &    26 & 1468 &  -3.91 &   0.08 & 0.2      &  0.0068 \\
mv5004044 &  719 &  3.8 & 21.0 &  9536 &    26 & 1770 &  -3.60 &   0.06 & $<$0.1   &  0.0089 \\
mv5004046 &  697 &  4.3 & 25.4 &  9012 &    25 & 1391 &  -3.76 &   0.07 & 0.1      &  0.0087 \\
mv5004051 &  696 &  3.7 & 29.2 &  9695 &    26 & 1769 &  -3.65 &   0.06 & $<$0.1   &  0.0104 \\
mv5004053 &  715 &  3.5 & 23.1 &  9603 &    26 & 1769 &  -3.57 &   0.06 & $<$0.1   &  0.0101 \\
mv5004056 &  753 &  3.4 & 20.7 &  9543 &    26 & 1806 &  -3.58 &   0.06 & $<$0.1   &  0.0083 \\
mv5004058 &  807 &  3.2 & 16.8 & 10378 &    27 & 1937 &  -3.65 &   0.06 & $<$0.1   &  0.0073 \\
mv5004061 &  882 &  3.0 & 24.0 &  8495 &    24 & 1681 &  -3.65 &   0.06 & $<$0.1   &  0.0056 \\
mv5004063 &  962 &  2.6 & 14.3 &  6967 &    21 & 1771 &  -3.64 &   0.06 & 1.2      &  0.0047 \\
mv5004066 & 1096 &  2.4 & 14.7 &  4796 &    17 & 1426 &  -3.82 &   0.07 & 1.2      &  0.0034 \\
mv5006001 & 1294 &  2.1 & 12.2 &  2373 &    11 &  965 &  -3.87 &   0.09 & 17.8     &  0.0021 \\

\hline\noalign{\smallskip}
\multicolumn{11}{c}{MRI 501 ms} \\
\hline\noalign{\smallskip}
mv5002051 & 5148 &  0.4 &  1.1 &    54 &     1 &  105 &  -4.54 &   0.35 & 82.3     &  0.0002 \\
mv5002061 & 4082 &  0.5 &  1.7 &   102 &     1 &  192 &  -4.70 &   0.27 & 92.6     &  0.0003 \\
mv5002069 & 3210 &  0.7 &  3.0 &   243 &     2 &  292 &  -4.26 &   0.19 & 5.8      &  0.0005 \\
mv5004001 & 2523 &  0.8 &  3.4 &   653 &     3 &  721 &  -3.76 &   0.10 & 11.6     &  0.0011 \\
mv5004005 & 2172 &  0.9 &  5.6 &  1052 &     5 &  837 &  -3.70 &   0.09 & 17.6     &  0.0013 \\
mv5004009 & 1858 &  1.1 &  6.7 &  1526 &     7 & 1034 &  -3.69 &   0.08 & 0.1      &  0.0017 \\
mv5006016 & 1886 &  0.9 &  7.3 &  1288 &     6 &  991 &  -3.59 &   0.08 & 9.0      &  0.0015 \\
mv5006020 & 2219 &  0.7 &  4.3 &  1028 &     5 &  954 &  -3.72 &   0.09 & 4.2      &  0.0013 \\
mv5006024 & 2643 &  0.7 &  3.2 &   400 &     3 &  411 &  -3.81 &   0.14 & 39.6     &  0.0007 \\
mv5006030 & 3345 &  0.7 &  2.8 &   166 &     2 &  184 &  -4.04 &   0.22 & 63.0     &  0.0003 \\
mv5006037 & 4187 &  0.5 &  1.6 &    81 &     1 &  120 &  -4.85 &   0.35 & 37.5     &  0.0002 \\

\end{longtable}
\end{SmallLongTable}
\clearpage

\begin{table}
  \renewcommand\baselinestretch{1}
  \footnotesize
  \begin{center}
    \caption[Power-law fit summary]{Summary of single power-law fits
      to individual HRIVIS and MRI flux distributions listed in
      Table~\ref{tab:fits}.  Table columns: $\overline{F_{\lambda}}$,
      mean minimum and maximum particle fluxes; $\overline{N}$, mean
      number of particles; $\alpha$, minimum, maximum, mean, and
      standard deviation of the flux distribution power-law slopes;
      $\overline{P_{KS}}$, mean Kolmogorov-Smirnov probability that
      the distributions are power-laws; $\overline{F_T / F_{coma}}$,
      mean fraction of the total coma flux attributable to point
      sources with $F_{\lambda} > F_{\lambda, {\rm min}}$ in MRI
      full-frame data, excluding regions near the
      nucleus.}  \label{tab:fit-summary}
    \begin{tabular}{l rr@{\extracolsep{6pt}} r@{\extracolsep{0pt}}r r rrrr rr}
      \hline\noalign{\smallskip}
      
      Instrument &
      \multicolumn{2}{c}{$\Delta$} &
      \multicolumn{2}{c}{$\overline{F_{\lambda}}$} &
      $\overline{N}$ &
      \multicolumn{4}{c}{$\alpha$} &
      $\overline{P_{KS}}$ &
      $\overline{F_T / F_{coma}}$
      \\\cline{2-3}\cline{4-5}\cline{7-10}
      &
      min &
      max &
      min &
      max &
      &
      \multicolumn{1}{c}{min} &
      \multicolumn{1}{c}{max} &
      \multicolumn{1}{c}{mean} &
      \multicolumn{1}{c}{$\sigma$} &
      
      \\
      &
      \multicolumn{2}{c}{(km)} &
      \multicolumn{2}{p{15mm}}{{(\scriptsize $10^{-12}$ W~m$^{-2}$~\micron$^{-1}$)}} &
      &
      &
      &
      &
      &
      (\%) &

      \\
      \hline\noalign{\smallskip}

      HRI           &  694 &  915 &  1.3 & 30.0 &  545 &  -3.08 &  -2.67 &  -2.87 &   0.14 & 0.125 & \nodata \\
      MRI sub-frame &  696 &  964 &  3.6 & 14.2 &  127 &  -4.31 &  -3.22 &  -3.70 &   0.35 & 0.341 & \nodata \\
      MRI 41 ms     &  694 & 1006 &  5.5 & 20.2 &  245 &  -4.39 &  -3.59 &  -3.90 &   0.22 & 0.384 & 0.0023 \\
      MRI 121 ms    &  696 & 1303 &  3.1 & 17.0 & 1489 &  -3.95 &  -3.57 &  -3.76 &   0.14 & 0.021 & 0.0059 \\
      MRI 501 ms    & 1858 & 5148 &  0.7 &  3.5 &  531 &  -4.85 &  -3.59 &  -4.06 &   0.43 & 0.330 & 0.0008 \\
      MRI (distant)$^{{\rm a}}$ & 2172 & 2643 &  9.0 & 46.9 &  731 &  -3.81 &  -3.70 &  -3.75 &   0.04 & 0.182 & 0.0011 \\
      \hline
    \end{tabular}
  \end{center}
  $^{{\rm a}}$ Averages of the the four 501 ms MRI images with
  $\Delta>2000$~km and $N>300$ particles, with which we use to define
  the fraction of the coma flux attributable to large particles.  For
  this row only, $F_\lambda$ has been corrected to a distance of
  $\Delta=700$~km before averaging.
\end{table}

\begin{table}
  \scriptsize
  \caption[Extrapolation]{Flux range (\fmin, \fmax), radius range
    ($a_{icy}$), total number ($N$), flux ($F_{\lambda,T}$), fraction
    of coma flux in particles ($F_T / F_{coma}$), cross section
    ($\sigma_{icy}$), mass ($M_{icy}$), and water production rate
    ($Q(\water)$) for the large particles measured at
    $\Delta=2100-2600$.  The radius, cross section, mass, and water
    production rate are computed using an icy composition with
    Europa's photometric parameters (at $\phi=86$\degr) and a density
    of 1.0~g~cm$^{-3}$.  We also extrapolate these quantities down to
    $10^{-14}$ W~m$^{-2}$~\micron$^{-1}$, and to our total coma
    measurement for a 20.6~km aperture at $\Delta=5148$~km.
    Uncertainties are derived from the power-law slope uncertainties
    ($\pm0.02$) and are symmetric in log space.
    \label{tab:extrapolate}
  }

  \begin{center}
  \begin{tabular}{ll rr *{7}{r@{}l}}
    \hline\noalign{\smallskip}
    \multicolumn{1}{c}{\fmin$^{\rm a}$} &
    \multicolumn{1}{c}{\fmax$^{\rm a}$} &
    \multicolumn{2}{c}{$a_{icy}$} &
    \multicolumn{2}{c}{$\log_{10}N$} &
    \multicolumn{2}{c}{$\log_{10}F_{\lambda,T}$$^{\rm a}$} &
    \multicolumn{2}{c}{$\frac{F_T}{F_{coma}}$} &
    \multicolumn{2}{c}{$\log_{10}\sigma_{icy}$} &
    \multicolumn{2}{c}{$\log_{10}M_{icy}$} &
    \multicolumn{2}{c}{$\log_{10}Q_{max}$}
    \\
    \multicolumn{2}{c}{$\times10^{-12}$}
    \\
    \multicolumn{2}{c}{(W~m$^{-2}$~\micron$^{-1}$)} &
    \multicolumn{2}{c}{(cm)} &
    & &
    \multicolumn{2}{c}{(W~m$^{-2}$~\micron$^{-1}$)} &
    \multicolumn{2}{c}{(\%)} &
    \multicolumn{2}{c}{(cm$^2$)} &
    \multicolumn{2}{c}{(g)} &
    \multicolumn{2}{c}{(s$^{-1}$)} \\
    \hline\noalign{\smallskip}

    \multicolumn{16}{c}{$\Delta\approx2400$~km, full-frame (excluding masked regions)} \\
    \hline\noalign{\smallskip}

    9.00 $^{\rm b}$ &45 $^{\rm b}$ &7.4 &16.5 &2.86 &                    $^{\rm b}$ &-8.06 &                   $^{\rm b,c}$ &0.11 &                $^{\rm b}$ &5.22 &                    &9.07 &                    &22.56 &                   &\\ 
    1.00 &45 &2.5 &16.5 &4.90 & $\pm$0.02                    &-6.86 & $\pm$0.02                   &1.8 & $\pm0.1$              &6.42 & $\pm$0.02                    &9.92 & $\pm$0.01                    &23.77 & $\pm$0.02                   &\\ 
    0.10 $^{\rm e}$&45 &0.8 &16.5 &6.76 & $\pm$0.04                    &-5.97 & $\pm$0.03                   &13.8 & $\pm1.1$             &7.32 & $\pm$0.03                    &10.39 & $\pm$0.03                   &24.66 & $\pm$0.03                   &\\ 
    0.01 &45 &0.2 &16.5 &8.61 & $\pm$0.06                    &-5.11 & $\pm$0.05                   &98.7 & $^{+12.8}_{-11.3}$           &8.17 & $\pm$0.05                    &10.79 & $\pm$0.04                   &25.52 & $\pm$0.05                   &\\ 

    \hline\noalign{\smallskip}
    \multicolumn{16}{c}{$\Delta=5148$ km, 20.6~km aperture} \\
    \hline\noalign{\smallskip}

    1.00 &45 &2.5 &16.5 &6.28 & $\pm$0.03                    &-6.69 & $\pm$0.02                   &1.8 & $\pm0.1$              $^{\rm d}$ &6.59 & $\pm$0.02                    &11.30 & $\pm$0.03                   &23.94 & $\pm$0.02                   &\\ 
    0.10 $^{\rm e}$&45 &0.8 &16.5 &9.02 & $\pm$0.07                    &-5.79 & $\pm$0.03                   &13.8 & $\pm1.1$ $^{\rm d}$           &7.49 & $\pm$0.03                    &12.66 & $\pm$0.06                   &24.83 & $\pm$0.03                   &\\ 
    \hline\noalign{\smallskip}
  \end{tabular}
  \end{center}

  $^{\rm a}$ Fluxes have been corrected to a distance of 700~km.

  $^{\rm b}$ Measured values.

  $^{\rm c}$ The instrumental and calibration uncertainties are
  $2\times 10^{-12}$ W~m$^{-2}$~\micron$^{-1}${} and 10\%, respectively.

  $^{\rm d}$ Assumed value.

  $^{\rm e}$ Our preferred lower flux limit is between 0.1 and
  1.0$\times10^{-12}$ W~m$^{-2}$~\micron$^{-1}$.
\end{table}

\begin{table}
  \small
  \caption[Water production]{Summary of comet Hartley~2 water
    production rates observed near perihelion, and derived in this
    study. \label{tab:qwater}}

  \begin{tabular}{l c r@{$\,\pm\,$}l l}
    \hline\noalign{\smallskip}
    \multicolumn{1}{c}{Date(s)} &
    \multicolumn{1}{c}{Aper.$^{\rm a}$} &
    \multicolumn{2}{c}{$Q(\water)$} &
    \multicolumn{1}{c}{Notes}
    \\
    \multicolumn{1}{c}{(UT, 2010)} &
    \multicolumn{1}{c}{(km)} &
    \multicolumn{2}{c}{($10^{27}$ molec~s$^{-1}$)} &
    \\
    \hline\noalign{\smallskip}
    18 Oct     & $9.1\times 10^{5}$ & 8.70 & 0.38 & \citealt{combi11-h2} \\
    19 Oct     & 19              & 3.45 & 0.04 & \citealt{mumma11-h2} \\
    22 Oct     & 19              & 6.78 & 0.26 & \citealt{mumma11-h2} \\
    23--31 Oct & $1.6\times 10^{4}$ &
    \multicolumn{1}{r}{$\lesssim$2.\phantom{$\pm$}} & & Large particles only, \citealt{harmon11} + this work \\
    31 Oct$^{\rm b}$ & 3800        & 11.6 & 0.7  & \citealt{knight12-thisissue} \\
    1 Nov     & $1.1\times 10^{6}$ & 6.38 & 0.12 & \citealt{combi11-h2} \\
    2 Nov     & 300             & \multicolumn{1}{r}{10.\phantom{$\pm$}} & & \citealt{ahearn11} \\
    4 Nov     & 22              & 11.45 & 0.65 & \citealt{dellorusso11}  \\
    4 Nov     & 20.6            &
    \multicolumn{1}{r}{$\lesssim$0.004\phantom{$\pm$}} & & Large particles only, Table~\ref{tab:extrapolate} \\

    \hline\noalign{\smallskip}

  \end{tabular}

  \begin{footnotesize}
    $^{\rm a}$ Slit half-width or aperture radius.  We have assumed
    the 0.43\arcsec{} slit for the \citet{mumma11-h2} observations,
    and a 3~pixel radius for the \citet{combi11-h2} observations.

    $^{\rm b}$ Mean and standard deviation of 5 measurements taken on
    31 Oct.
  \end{footnotesize}
\end{table}

\clearpage
\cfoot{Kelley et al.\ 2015, Icarus, in press}
\thispagestyle{fancy}
\setcounter{page}{1}
{\center
  \Large Erratum to ``A distribution of large particles in the coma of
  Comet 103P/Hartley 2'': [Icarus 222, 634--652 (2013)]\par\vskip18pt

  \normalsize Michael S. P. Kelley\textsuperscript{a,*},
  Don J. Lindler\textsuperscript{b},
  Dennis Bodewits\textsuperscript{a},
  Michael F. A'Hearn\textsuperscript{a},
  Carey M. Lisse\textsuperscript{c},
  Ludmilla Kolokolova\textsuperscript{a},
  Jochen Kissel\textsuperscript{d,1},
  Brendan Hermalyn\textsuperscript{e}
  \par\vskip10pt

  \footnotesize\itshape
  \textsuperscript{a}{Department of Astronomy, University of Maryland, College
  Park, MD 20742-2421, USA}\\
  \textsuperscript{b}{Sigma Space Corporation, 4600 Forbes Boulevard,
  Lanham, MD 20706, USA}\\
  \textsuperscript{c}{Johns Hopkins University--Applied Physics Laboratory,
  11100 Johns Hopkins Road, Laurel, MD 20723, USA}\\
  \textsuperscript{d}{Max-Planck-Institut f\"ur Sonnensystemforschung,
  Max-Planck-Str. 2, 37191 Katlenburg-Lindau, Germany}\\
  \textsuperscript{e}{NASA Ames Research Center / SETI Institute, MS
  245-3 BLDG245, Moffett Field, CA 94035}
  \par\vskip36pt
  \hrule
}

\footnotetext{retired}


\setcounter{section}{0}
\section{Introduction}

Comet 103P/Hartley~2 has a high water production rate to surface area
ratio, suggesting the nucleus is nearly 100\% active.  In contrast,
images of the nucleus from the \textit{Deep Impact Flyby} spacecraft
show strong localized activity, with an inner-coma enriched in \coo{}
gas and water-ice grains.  Rather than being produced solely by water
ice sublimation at the nucleus, the hyperactivity of Hartley 2 may be
due to water-ice sublimation in the coma.  However, the contribution
of coma grains is poorly constrained, leaving the icy-grain hypothesis
unproven \citep{ahearn11, protopapa14}.

Images from the \textit{Deep Impact Flyby} spacecraft show thousands
of point sources surrounding the nucleus of comet Hartley 2
\citep{ahearn11}.  The point sources are individual particles
ejected by the comet.  We measured their brightnesses and summarized
their sizes, total mass, and spatial distribution
\citep{kelley13-dixi}.  The photometric properties (albedo, phase
function) of the particles are unknown, therefore we adopted two
models as approximate limiting cases: 1) bright, icy particles with
photometric properties similar to the Jovian satellite Europa; and 2)
dark, nucleus-like particles with properties similar to the nucleus of
Hartley 2.  For the bright, icy case, we reported the largest particle
had a radius of 20~cm and the total population mass within 21~km of
the nucleus was up to 3 to 10\% of the nucleus mass (assuming a
density of 1~g~cm$^{-3}$).  For the dark, nucleus-like case, the
largest particle was 2~m in radius and the total population up to 230
to 700\% of the nucleus mass (0.3~g~cm$^{-3}$).  Based on this mass
calculation, we ruled out the nucleus-like case and favored the icy
case.  Moreover, the icy case produces particles too small to account
for the comet's hyperactivity.

We have found three errors in our calculations that affected our
radius and mass estimates.  We regret the errors, as they have
significant consequences in our analysis of the particles.  We provide
updated calculations and interpretations in
Sections~\ref{sec:corrections} and \ref{sec:interpretation}.  In
addition, we provide a few minor clarifications and corrections to our
original paper in Section~\ref{sec:other}.

\section{Radius and mass corrections}\label{sec:corrections}

We have found an error that affects the computed particle radii (and
dependent quantities), and two additional errors that affect
population masses.  First, Eq.~4 of \citet{kelley13-dixi} is missing a
factor of $\pi$ in the denominator.  The correct equation is
\setcounter{equation}{3}
\begin{equation}
  \label{eq:flux}
  F_\lambda = \frac{A_p\Phi(\theta)\sigma S_{\lambda,\sun}}{\pi r_h^2 \Delta^2}
\end{equation}
where $F_\lambda$ is the particle flux density, $A_p$ is the geometric
albedo, $\Phi(\theta)$ is the phase function evaluated at the phase
angle $\theta$, $\sigma$ is the cross-sectional area of the particle,
$S_{\lambda,\odot}$ is the solar flux density at 1~AU, $r_h$ is the
heliocentric distance of the particle, and $\Delta$ is the
spacecraft-particle distance.  To account for this factor of $\pi$,
all parameters derived from particle fluxes must be scaled as follows:
radius by a factor of $\sqrt{\pi}$, cross section and water production
rate by a factor of $\pi$, and mass by a factor of $\pi^{3/2}$.

Second, our computations of total population mass (Table 4)
erroneously used 1000~g\,cm$^{-3}$, rather than the 1~g\,cm$^{-3}$
quoted in the text.  In addition, our analytical formula for
integrating the total mass of a population of grains given a power-law
flux distribution was missing a factor of 2.  The formula was not
given in the paper, but for completeness the corrected formula is
\begin{equation*}
  \label{eq:integral}
  M = \frac{8 \pi \rho}{3} N_0 F_1^{\alpha - 1} \int_{a_{min}}^{a_{max}}{a^{2 \alpha + 2}da},
\end{equation*}
where $M$ is the total population mass, $\rho$ is the particle mass
density, $N_0$ is the flux distribution normalization constant, $F_1$
is the flux density from a 1-cm radius particle (via
Eq.~\ref{eq:flux}), $a$ is the particle radius, $a_{min}$, $a_{max}$
are the limits of the integration (corresponding to the estimated flux
density limits), and $\alpha$ is the power-law index of the
differential flux distribution (Eq.~3).  Thus, total
\textit{population} masses must be scaled by a factor of $1/500$, in
addition to the $\pi^{3/2}$ scale factor from above (the total scale
factor is 0.011).  Other quantities that depend on \textit{individual}
particle mass (e.g., $\beta$ in Section~6.2, $\alpha$ and $v_{ej}$ in
Section~6.3) are unaffected.

We present a revised Table~4, with corrected radii, cross sections,
mass, and maximum water production rates.  In addition, all
calculations are now based on the normalization constant $N_0$
(Eq.~6), whereas previously some calculations were normalized via the
total observed particle flux.  This change increases most flux-based
quantities by 13\%.  Overall, our analysis is significantly affected
by the revised Table~4.

The solid water-ice case (see Section 5 of \citealt{kelley13-dixi} for
definitions of our icy and dusty cases) produces particle estimates up
to 30~cm in radius with an estimated population mass (within a 20.6~km
aperture) of $0.2-6\times10^{10}$~g, or up to 0.02\% of the nucleus
mass.  This mass is two orders of magnitude lower than the $\sim2$\%
of the nucleus that was lost in 2010 \citep{thomas13-hartley2}.
Porous ice particles are no longer necessary to reduce the estimated
population mass below the orbital mass loss of the comet.  The total
water production rate estimate is
$Q_{max}=0.3-2\times10^{25}$~s$^{-1}$, or $0.03-0.2$\% of the comet's
total water production rate, insufficient to account for the comet's
hyperactivity.

To convert from the icy case to the dusty case multiply radii by 12.6,
cross sections by 158, masses by 597, and $Q_{max}$ by 1007.  The
dusty particle case produces particles up to 4~m in radius and a total
population mass of $0.1-3\times10^{13}$~g, or 0.6--14\% of the mass of
the nucleus, potentially exceeding the estimated orbital mass-loss
rate of the comet.  If the dusty particles behave like mini-comet
nuclei, their total water production rate may be as large as
$Q_{max}=0.3-2\times10^{28}$~s$^{-1}$, or 30--200\% of that of the
comet during the encounter.  Our updated cross sections,
0.2--2~km$^2$, are comparable to and slightly exceed the cross section
measured by radar observations \citep[0.89~km$^2$;][]{harmon11}.

The new radius estimates produce particles up to 8~m diameter.  Such
particles should be resolved in the High-Resolution Instrument (HRI)
visible images, which has a reconstructed resolution of approximately
3~m at closest approach.  However, our original examination of the
data revealed no resolved sources, suggesting a maximum size of
approximately 3~m.  The smaller size could be accommodated through an
increase in the model albedo (from 0.049 to 0.31) or by reducing the
phase function coefficient (from 0.046 to 0.023 mag per degree; see
Section 5 of \citealt{kelley13-dixi}).  Both cases reduce the total
population mass to $<$1\% of the nucleus mass and maximum water
production rate to 40\% of the total comet production rate.  We save a
more thorough search for resolved HRI particles for future work.

\section{Interpretation}\label{sec:interpretation}
Given the above revisions, the dusty case may solve the apparent
hyperactivity of the comet.  However, if these particles are long
lived (i.e., not fragmenting) and moving away with radial velocities
of order 0.1~m~s$^{-1}$ \citep{hermalyn13-dixi}, then their lifetimes
in a 21~km radius aperture are of order 60 hours, implying a
significantly large mass-loss rate of $\sim10^3$~kg~s$^{-1}$ for the
assumed particle density of 0.3~g~cm$^{-3}$.  This mass-loss rate must
be increased by an order of magnitude if we instead consider the radar
observed velocity dispersion of 4~m~s$^{-1}$ \citep{harmon11}.  These
estimates significantly exceed the comet's total water and carbon
dioxide production rates \citep[$\sim300$~kg~s$^{-1}$ and
$\sim160$~kg~s$^{-1}$, respectively;][]{ahearn11} near the time of the
flyby.  Therefore, understanding the large particle dynamics is
critical to determining the erosion rate of the comet, and if the
dusty case remains valid.  In parallel, a definitive upper limit to
particle sizes should be derived from a more detailed search for
resolved particles in HRI images.

We have outlined two dramatically different scenarios for the physical
properties of the large particles of Hartley~2.  Intermediate sets of
photometric and compositional properties are possible, and such cases
cannot be ruled out.  We do consider the dusty case (low albedo,
0.3~g~cm$^{-3}$) to be less likely due to the large and massive
particles implied, but cannot confidently rule it out at this time.
However, note that higher albedos or different phase functions may be
employed to produce a more physically consistent picture for the large
particles.

\section{Other corrections}\label{sec:other}

Equation 5 is correctly referred to as the differential size
distribution, but elsewhere the text commonly omits the
``differential'' adjective.  We reviewed our comparisons to other
estimates of the dust size distribution, and the nomenclature used in
the literature is not very specific, occasionally labeling
differential size distributions as size distributions (just as we
did).  Our best understanding of the investigations listed in the text
is that most indeed report the differential size distribution.  Our
conclusion, that the differential size distribution slope, $-4.7$, is
steeper than other estimates, remains valid.

The normalization constant $N_0$ is not needed in Eq.~6, it is already
included in Eq.~5.  The corrected Eq.~6 is
\setcounter{equation}{5}
\begin{equation}
  N = \int_{F_{\lambda,{\rm min}}}^{F_{\lambda,{\rm max}}}\frac{dn}{dF}dF.
\end{equation}
The calculations for the paper used the above formula.

The nucleus mass listed in Section~7 was expressed in units of kg, but
reported as g.  The nucleus mass is $M_N=2.4\times10^{14}$~g
\citep{thomas13-hartley2}.

Finally, the units of flux density in Section 4.2 should be
W~m$^{-2}$~\micron$^{-1}$, and not W~cm$^{-2}$~\micron$^{-1}$ as
stated.

\section*{Acknowledgments}

We thank Katherine Kretke for identifying the mass discrepancy in
Table 4, and Michael Belton for motivating other clarifying comments.

This work was supported by NASA (USA) through the Planetary Mission
Data Analysis Program contract NNX12AQ64G to the University of
Maryland.

\setcounter{table}{3}
\begin{table}
  \scriptsize
  \caption[Extrapolation]{Flux range ($F_{\lambda,min}$,
    $F_{\lambda,max}$), radius range ($a_{icy}$), total number
    ($N$), total flux ($F_{\lambda,T}$), fraction 
    of coma flux in particles ($F_T / F_{coma}$), cross section
    ($\sigma_{icy}$), mass ($M_{icy}$), and maximum water production rate
    ($Q_{max}(\water)$) for the large particles measured at
    $\Delta=2100-2600$.  The radius, cross section, mass, and water
    production rate are computed using an icy composition with
    Europa's photometric parameters (at $\phi=86$\degr) and a density
    of 1.0~g~cm$^{-3}$.  We also extrapolate these quantities down to
    $10^{-14}$ W~m$^{-2}$~\micron$^{-1}$, and to our total coma
    measurement for a 20.6~km aperture at $\Delta=5148$~km.
    Uncertainties are derived from the power-law slope uncertainties
    ($\pm0.02$) and are symmetric in log space.
    \label{tab:extrapolate}
  }

  \begin{center}
  \begin{tabular}{ll rr *{7}{r@{}l}}
    \hline\noalign{\smallskip}
    \multicolumn{1}{c}{{$F_{\lambda,min}$}$^{\rm a}$} &
    \multicolumn{1}{c}{{$F_{\lambda,max}$}$^{\rm a}$} &
    \multicolumn{2}{c}{$a_{icy}$} &
    \multicolumn{2}{c}{$\log_{10}N$} &
    \multicolumn{2}{c}{$\log_{10}F_{\lambda,T}$$^{\rm a}$} &
    \multicolumn{2}{c}{$\frac{F_T}{F_{coma}}$} &
    \multicolumn{2}{c}{$\log_{10}\sigma_{icy}$} &
    \multicolumn{2}{c}{$\log_{10}M_{icy}$} &
    \multicolumn{2}{c}{$\log_{10}Q_{max}$}
    \\
    \multicolumn{2}{c}{$\times10^{-12}$}
    \\
    \multicolumn{2}{c}{(W~m$^{-2}$~\micron$^{-1}$)} &
    \multicolumn{2}{c}{(cm)} &
    & &
    \multicolumn{2}{c}{(W~m$^{-2}$~\micron$^{-1}$)} &
    \multicolumn{2}{c}{(\%)} &
    \multicolumn{2}{c}{(cm$^2$)} &
    \multicolumn{2}{c}{(g)} &
    \multicolumn{2}{c}{(s$^{-1}$)} \\
    \hline\noalign{\smallskip}

    \multicolumn{16}{c}{$\Delta\approx2400$~km, full-frame (excluding masked regions)} \\
    \hline\noalign{\smallskip}

    9.00 $^{\rm b}$ &45 $^{\rm b}$ &13.1 &29.3 &2.86 & $\pm$0.00 $^{\rm b}$ &-8.01 & $\pm$0.00 $^{\rm c}$ &0.1 & $\pm$0.0              &5.77 & $\pm$0.00                    &7.12 & $\pm$0.00                    &23.11 & $\pm$0.00                   &\\ 
    1.00 &45 &4.4 &29.3 &4.90 & $\pm$0.02                    &-6.80 & $\pm$0.01                   &2.0 & $\pm$0.1              &6.97 & $\pm$0.01                    &7.96 & $\pm$0.01                    &24.32 & $\pm$0.01                   &\\ 
    0.10 &45 &1.4 &29.3 &6.76 & $\pm$0.04                    &-5.91 & $\pm$0.03                   &15.5 & $\pm$1.2             &7.86 & $\pm$0.03                    &8.44 & $\pm$0.03                    &25.21 & $\pm$0.03                   &\\ 
    0.01 &45 &0.4 &29.3 &8.61 & $\pm$0.06                    &-5.06 & $\pm$0.05                   &111 & $\pm$13               &8.72 & $\pm$0.05                    &8.84 & $\pm$0.04                    &26.06 & $\pm$0.05                   &\\ 

    \hline\noalign{\smallskip}
    \multicolumn{16}{c}{$\Delta=5148$ km, 20.6~km aperture} \\
    \hline\noalign{\smallskip}

    1.00 &45 &4.4 &29.3 &6.33 & $\pm$0.03                    &-6.63 & $\pm$0.01                   &2.0 & $\pm$0.1 $^{\rm d}$              &7.14 & $\pm$0.01                    &9.39 & $\pm$0.03                    &24.49 & $\pm$0.01                   &\\ 
    0.10 &45 &1.4 &29.3 &9.08 & $\pm$0.07                    &-5.74 & $\pm$0.03                   &15.5 & $\pm$1.2  $^{\rm d}$            &8.04 & $\pm$0.03                    &10.76 & $\pm$0.06                   &25.38 & $\pm$0.03                   &\\ 

    \hline\noalign{\smallskip}
  \end{tabular}
  \end{center}

  $^{\rm a}$ Fluxes have been corrected to a distance of 700~km.

  $^{\rm b}$ Measured values.

  $^{\rm c}$ The instrumental and calibration uncertainties are
  $2\times 10^{-12}$ W~m$^{-2}$~\micron$^{-1}${} and 10\%, respectively.

  $^{\rm d}$ Assumed value.

  $^{\rm e}$ Our preferred lower flux limit is between 0.1 and
  1.0$\times10^{-12}$ W~m$^{-2}$~\micron$^{-1}$.
\end{table}
\end{document}